\documentclass[11pt]{article}
\usepackage[utf8]{inputenc}
\usepackage{authblk}
\usepackage{dirtytalk, float}
\usepackage[top=.8in, bottom=0.75in, left=0.75in, right=0.75in]{geometry}
\usepackage{epsfig, color}
\usepackage{amsfonts, amsmath, setspace}
\usepackage{mathtools, array, booktabs}
\usepackage{amssymb, cite, soul, url}
\usepackage[normalem]{ulem}
\usepackage{scalerel}
\usepackage{graphicx, float, caption, subcaption} 
\title{Investigating Buoyant Plume Dynamics Induced by Localized Fire-Simulated Heating over Plant Canopies Using LES}
\author[1,2]{Ajinkya Desai}
\author[2]{Antonio Quim Cervantes}
\author[2,3]{Tirtha Banerjee}
\affil[1]{Lawrence Livermore National Laboratory\footnote{Release Number: LLNL-JRNL-2012493}, Livermore, CA 94550}
\affil[2]{Department of Civil and Environmental Engineering, University of California, Irvine, CA 92697}
\affil[3]{Department of Earth System Science, University of California, Irvine, CA 92697}\date{}

\begin{document}
\maketitle
\begin{abstract}

The interaction of a buoyant plume with a plant canopy results in turbulent flow features distinct from those in a grassland environment. In this work, we model the turbulence dynamics of a buoyant plume in a homogeneous plant canopy with a crosswind using large-eddy simulations. As the plume interacts with the crosswind, we observe increased vorticity at the windward edge and tilted hair-pin-like vortical structures on the leeward side. Strong rotational cores, representing counter-rotating vortex pairs (CVPs), form as the flow twists and spirals into the leeward side of the buoyancy source from either side. Flow patterns aloft exhibit helical motions as the CVPs aloft propagate downstream, trailing the plume. We also simulate a no-canopy environment to facilitate comparison. The plume tilts less steeply near the source in the canopy case due to the canopy drag and its leeward side is marked by flow recirculation near the canopy top, which obstructs the upstream flow as it approaches. Moreover, the plume transition from the rise phase to the bent-over phase is delayed due to the canopy's aerodynamic effects and the oscillatory behavior of the far-field mean plume centerline is more damped. Additionally, in the canopy environment, there is downward momentum transfer primarily via ejections above the canopy and sweeps within the canopy space, upstream of the plume centerline. On the leeward side, counter-gradient motions play a significant role in transferring momentum away from the buoyancy source, with outward interactions being most dominant. Contrarily, in the no-canopy environment, counter-gradient motions near the surface are flanked upstream by an ejection-dominated region and downstream by a sweep-dominated region. Insights into the distinct plume behavior in canopy vs. no-canopy environments are vital for comparing with experiments and refining fire behavior or plume rise models.
 \\

\end{abstract}

\section{Introduction}\label{sect_intro}
Turbulent structures germinate from the interaction of a buoyant plume, such as that generated by a wildland fire, with an ambient cross-wind. Although the physical implications of such structures on wildland fire behavior and spread characteristics have only recently gathered much attention, plume--cross-wind interactions have long been of interest to fluid mechanics researchers with laboratory-scale experiments representing a major source of understanding. 
 Early laboratory experiments by Fric and Roshko \cite{fric1994vortical} on the interaction of a (non-thermal) jet issued from a stationary source with cross-flow, identified the presence of near-source shear-layer vortices and downstream counter-rotating vortex pairs
(CVPs). For a similar configuration, Cardenas et al. \cite{cardenas2007two} obtained two-dimensional maps of turbulent fluxes of tracers and momentum using simultaneous 2D-LIF and PIV measurements; CVPs were found to significantly influence the fluxes. Other laboratory experiments on jets-in-cross-flow, identifying CVPs and their potential formation mechanisms, such as the modulation of the near-source jet vorticity by the cross-wind, have been reviewed by Mahesh \cite{mahesh2013interaction}, while recent propagating-fire experiments in small-scale fuel beds \cite{finney2015role, desai2022investigating} have identified and quantified CVPs as key participants in fire spread.
Based on similarity criteria, laboratory studies have also attempted to develop power-law scalings characterizing plume trajectories under cross-flow in the near-source and far-field regions influenced by the ratio of the momentum flux to the buoyancy flux \cite{briggs1984plume, weil1988, davidson1994dimensionless, pun1999behaviour, robins2000water, lee2003, contini2011comparison, davidson1994dimensionless, huq1996laboratory, james2022particle}. Using flume experiments with a model vegetation canopy, Chung and Koseff \cite{chung2023interaction} investigated the influence of canopy-induced turbulence on buoyant plumes by varying the intensity of the buoyancy source. The plume-trajectory centerline for the case with maximum turbulence was found to oscillate significantly more than the case with the lowest turbulence, with the dominant frequency corresponding to that of the Kelvin-Helmholtz rollers in the mixing layer and higher relative plume buoyancy suppressing this influence. While lab-scale analyses have been useful in their own right, they lack the complexity of more realistic, spatially evolving plume behavior in the environment. Moreover, studies investigating buoyant-plume interactions with tall vegetative canopies in the presence of cross-flow are scarce. 

\textit{In-situ} measurements of wind velocity and temperature from prescribed burn experiments  \cite{heilman2023atmospheric} have contributed significantly to the understanding of fire-plume behavior at field scales under differing vegetative characteristics. Studies have attempted to quantify fire-plume-modulated local winds and turbulence-generation and transport mechanisms \cite{clements2007observing, clements2008first, heilman2015observations, heilman2017atmospheric, heilman2019observations, bebieva2020role, heilman2021observations, clements2019fireflux}, differentiate between these patterns during fires in grassland
and forested environments under differing ambient wind-forcing conditions \cite{desai2023features, heilman2021observations}, and track fire-modulated ramp--cliff structures in a forest canopy as they evolved temporally,
along with their constitutive frequencies \cite{desai2024investigating}. 
Observations regarding fire-modulated heat- and momentum-flux events at multiple heights \cite{heilman2021observations, heilman2021turbulent} were particularly noteworthy. Downdrafts of high horizontal momentum (sweeps) were found to dominate the intensity of momentum-flux events in forest canopy environments. However, updrafts of high horizontal momentum (outward interactions) were found to compete significantly with sweeps in their contribution to the momentum flux, while also showing a high occurrence frequency compared to no-fire conditions. Despite their usefulness, field measurements are limited by their spatial localization and logistical obstacles to
deploying multiple instruments while maintaining quality control \cite{wharton2023capturing}. Furthermore, 
prescribed burns are limited by seasonal restrictions and weather-related uncertainties, the respiratory and visibility hazards germinating from the accompanying smoke plumes, and escape risks \cite{baijnath2022historical, li2025temporal}; there remains a dearth of measured data during prescribed burns in forested regions \cite{heilman2021turbulent}. In addition, while the larger-scale sampling of plumes during active fires has gained momentum recently \cite{strobach2023isolating, carroll2024measuring, lareau2022fire}, with recent radar-based measurements providing evidence for the presence of CVPs at the fire-front \cite{lareau2022fire}, these observations also remain sparse.

Given the deficit in observational datasets vis-\'a-vis the necessity to develop reliable predictive models for active fire containment and planning prescribed burns, the majority of coupled fire--atmosphere modeling endeavors have been focused on predicting fire perimeters with a few specific observations. 
Mueller et al. \cite{mueller2021detailed} utilized the Wildland-urban interface Fire Dynamics Simulator (WFDS) to simulate a 4.25-hectare experimental surface fire in a pine forest. The study also aimed to evaluate the impact of canopy fuel structure on fire behavior. While fire spread rates showed reasonable agreement with experimentally observed values, local winds induced by the fire were underpredicted in its wake. Benik et al. \cite{benik2023analysis} utilized WRF-SFIRE to simulate fire-induced circulations near the perimeter during the FireFlux2 grassland fire experiment \cite{clements2019fireflux}. Models tracked observed fire-front locations and the vertical plume structure to high levels of accuracy; however, SFIRE relies on the Rothermel model for fire spread, which is based on laboratory experiments, and the study itself was limited to a grassland burn.  Shamsei et al. \cite{shamsaei2023coupled} attempted to recreate the 2018, complex-terrain Camp Fire using WRF-Fire. While fire propagation rate predictions showed better agreement compared to the semi-empirical FARSITE model \cite{finney1998farsite}, several sources of uncertainty were identified along with the need for systematic studies to demystify them in the context of fire--atmosphere feedback flow. Roberts et al. \cite{roberts2024sensitivityof} attributed shortcomings in capturing these processes to the sensitivity of sensible
heat fluxes during fire events to accurate estimates of fuel loading inputs.  In the face of the limited predictive capabilities of existing models for fire behavior, it is worth exploring model capabilities in studying process-level fire-induced flow phenomena by systematically varying the environmental conditions \cite{finney2012need, speer2022wildland}. Quantifying the near-field (close to the buoyancy source) fire-induced flow at high spatial resolution will facilitate improved parameterizations within the coarser-resolution predictive models. 

However, the complex system of equations governing fire physics within the atmospheric boundary layer (ABL) makes simulations computationally intensive \cite{grishin1996general, sullivan2009wildland, canfield2014numerical, chen2018numerical, ahmed2024simulations}. Low-complexity models provide a practical starting point due to their reduced computational demands and can be
progressively refined with added complexities and improved physical underpinnings to create reliable operational models.
Although some studies have simulated cross-wind interactions with traverse jets or buoyant plumes with differing source geometries using turbulence models such as DNS or LES in such low-complexity configurations \cite{lavelle1997buoyancy, thurston2013large, mahesh2013interaction, cintolesi2019turbulent, lappa2019highly, behera2023characterization, quan2024large, cao2022characteristics}, simulations incorporating forest canopies and their comparisons with no-canopy environments remain rare.
The scarcity of both observations and simulations in exploring buoyant-plume--canopy--cross-wind interactions makes it difficult to establish a solid theoretical framework for plume behavior modulated by such interactions.
In this study, we employ an LES model to explore interactions between a thermal plume emanating from a localized surface source with a prescribed heat flux in the presence and absence of a homogeneous forest canopy. The advantage of LES models lies in their ability to characterize the most energetic turbulent eddies accurately across space with high spatial resolution. The key research question we intend to address through this study is: how does the presence of a tall vegetative canopy alter the near-field coherent structures and far-field plume behavior as the plume interacts with a cross-wind? Secondly, we investigate the similarities between the current, no-flame, low-complexity configuration with those involving both static and propagating fires. We start by providing a brief description of the open-source solver and the setup of our simulations in Sect.~\ref{sect_setup}. This is followed by a set of results in Sect.~\ref{sect_results}, discussing physical insights into the plume flow features from the mean statistics of the velocity and temperature fields and the spatial heterogeneity in the turbulent momentum-flux events. Finally, we summarize our deductions in Sect.~\ref{sect_conc}, while motivating and outlining pathways for future research.

\section{Computational Setup}\label{sect_setup}

The Parallelized Large-Eddy 
Simulation Model (PALM) is utilized with its in-built plant-canopy module to study the plume--canopy interaction. PALM solves the non-hydrostatic, incompressible Navier-Stoke equations, filtered by the grid size, in Boussinesq-approximated form \cite{maronga2020overview}. The domain is discretized using a staggered, Arakawa C grid \cite{arakawa1977computational}, so that scalar quantities are resolved at the center of each grid cell, while velocity components are computed at the edges, i.e. half a grid width apart from the scalars. 
For the advection terms in the governing (prognostic) equations, the default settings are used for discretization, i.e. an upwind-biased 5th-order differencing scheme \cite{wicker2002time} in combination with a 3rd-order Runge–Kutta time-stepping scheme after Williamson \cite{williamson1980low}. PALM uses an iterative, multigrid scheme to solve the Poisson equation for the modified perturbation pressure \cite{maronga2015parallelized} to allow for non-cyclic boundary conditions, as discussed later in this section. For the sub-grid-scale (SGS) terms, a 1.5-order turbulence closure model after Deardorff \cite{deardorff1980stratocumulus} is used, where a prognostic equation is solved for the SGS turbulence kinetic energy in addition to a gradient-diffusion parameterization for the SGS turbulent heat and momentum fluxes. More details about the PALM model and architecture can be found in the literature \cite{maronga2015parallelized, maronga2020overview}.

\begin{figure}
    \centering
    \begin{tabular}{cc}
      (a) Inlet velocity profile   & (b) Domain description \\
  \includegraphics[scale=0.425, trim={0 0 0 20pt}]{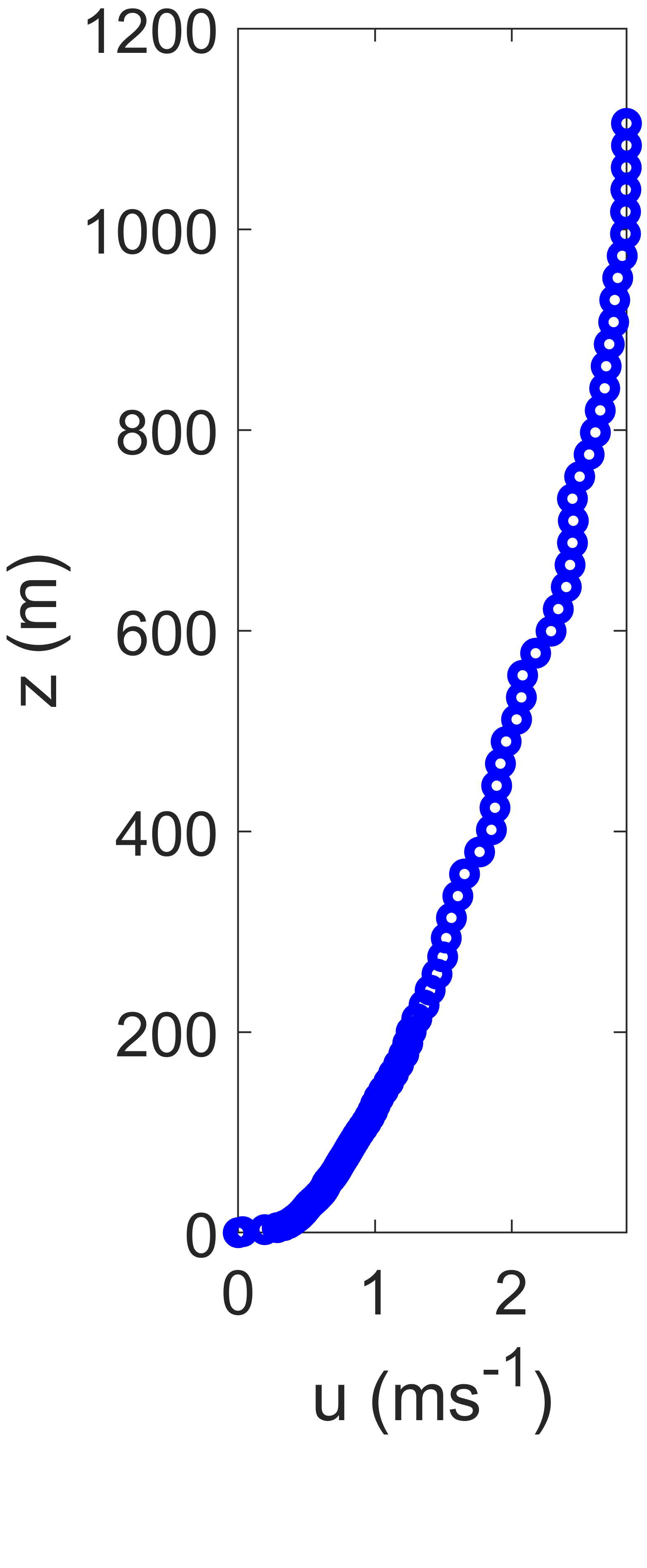}      & \hspace{15pt}\includegraphics[scale=0.425]{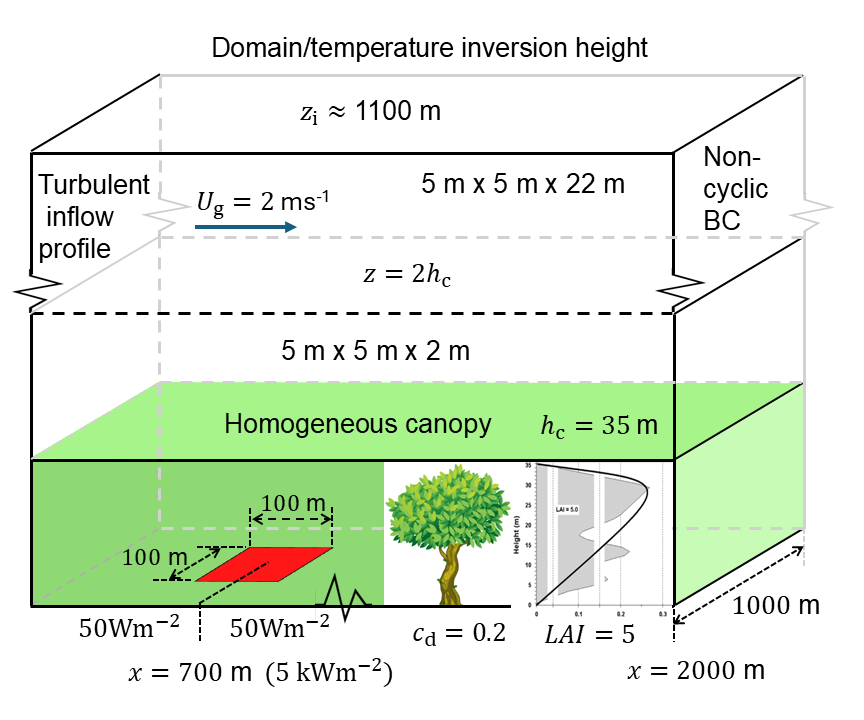}
     
    \end{tabular}
    \caption{A schematic of the computational setup for the simulations in a canopy setting. (a) Velocity profile at the inlet. (b) Domain description; kinks along the vertical axes represent regions with vertical grid stretching.}
    \label{fig_setup}
\end{figure}
Figure~\ref{fig_setup} encapsulates the computational setup for the LES run(s) in the canopy setting. The computational domain is 2\,km long in the streamwise ($x$) direction and 1\,km wide in the cross-stream ($y$) direction, with a homogeneous canopy up to a height ($z=h_\text{c}$) of 35\,m from the surface. The cell size for the structured grid is 5\,m$\times$5\,m$\times$2\,m up to $z=2h_\text{c}$, beyond which vertical grid stretching is employed using a stretch factor of 1.08, until the maximum cell size in the vertical direction is 22\,m. The higher resolution near the surface is aimed at capturing the complex turbulence patterns in the canopy sub-layer. The canopy has a drag coefficient ($c_\text{d}$) of 0.2 and leaf area index ($LAI$) of 5, with the corresponding leaf area density ($LAD$) profile taken from Dias-Junior et al. \cite{dias2015large}. The canopy elements are not explicitly modeled; rather, a volume-averaged effect of the porous canopy layer is considered, with the canopy drag force term in the momentum equation for the mean velocity, accounting for the aerodynamic effects of the canopy. This is given by $f_{\text{D}_{i}} = -c_\text{d}LADu_i\sqrt{u_{i}^{2}}$, where $u_i$ represents velocity components in the index notation. Daytime atmospheric conditions are simulated with a well-mixed boundary layer, i.e. a uniform initial profile of the potential temperature with height ($\theta(z) = \theta_\text{a} = 300$\,K) up to a temperature inversion height ($\delta$) of approximately 1.1\,km. Incoming solar radiation of 1015\,Wm\textsuperscript{-2} is assumed at the tree-tops \cite{patton2016atmospheric}, representative of early afternoon conditions. The ambient sensible heat flux at the forest surface was computed using a declining exponential function of the downward cumulative leaf area index ($F$) \cite{shaw1992large}: $Q(z) = Q(h_\text{c})e^{-\alpha F }$, where $\alpha$ is the extinction coefficient, taken as 0.6. Using, $Q(h_\text{c}) = 1015$\,Wm\textsuperscript{-2}, and $F=LAI=5$ at $z=0$, we get $Q(0)\approx 50$\,Wm\textsuperscript{-2}.

Fire plume presence is simulated by a
thermal plume arising from a localized region of size 100\,m\,$\times$\,100\,m to which a high surface heat flux is prescribed. The surface heat flux in this region is taken to be 100 times the ambient surface sensible heat flux ($Q=5000$\,Wm\textsuperscript{-2}). Fixing the heat source in space allows us to investigate the steady-state, micro-scale flow response to the plume. This approach allows us to bypass a combustion model that is highly sensitive to the accuracy of fuel-loading estimates to obtain the representative high sensible
heat fluxes \cite{roberts2024sensitivityof}. The reduced complexity allows us to quantify the effect of plume–wind–vegetation interactions
separately from the flow–combustion feedback \cite{lareau2024observations}. Previous
studies on buoyant plumes interacting with wind shear, reviewed by Lareau et al. \cite{lareau2022fire}, have demonstrated
the promise of this setup by capturing some characteristic flow features such as CVPs and shed vortex
columns. Non-cyclic horizontal boundary conditions (BC) are used for the inlet and outlet, whereas cyclic BC are used for the lateral (cross-stream) faces of the domain. The ambient wind follows a turbulent, logarithmic profile at the inlet. This turbulent inflow profile (Fig.~\ref{fig_setup}(a)) is achieved via a 48-hour precursor run in a smaller domain without a plant canopy, i.e. a turbulence spin-up run, with the geostrophic wind velocity ($U_\text{g}$) initialized to 2\,ms\textsuperscript{-1}. A technique called the turbulence-recycling method, described by Lund et al. \cite{lund1998generation}, is utilized for the main run comprising the larger domain, wherein turbulence from a cross-stream ($XY$) plane at $x=220$\,m is cycled back into the inlet for 2 h of simulation time. This is followed by several restart runs in which artificial momentum perturbations are periodically introduced at the inlet to facilitate atmospheric turbulence generation over a shorter fetch \cite{kumar2024impact}. The total simulation time is 9\,h. One-hour mean statistics ($\overline{u}_{i}, \overline{\theta}$) are computed via a block averaging over the last hour of the simulation ($t=8$--9\,h) and turbulent fluctuations are computed as deviations from the 1-h mean quantities: $u_i' = u_i - \overline{u}_{i}$ and $\theta' = \theta - \overline{\theta}$.

\section{Results}\label{sect_results}

\subsection{Mean velocity and temperature profiles}
\begin{figure}
    \centering
    \begin{tabular}{cc}
    (a) Canopy & (b) No-canopy\\
     \includegraphics[scale=0.35]{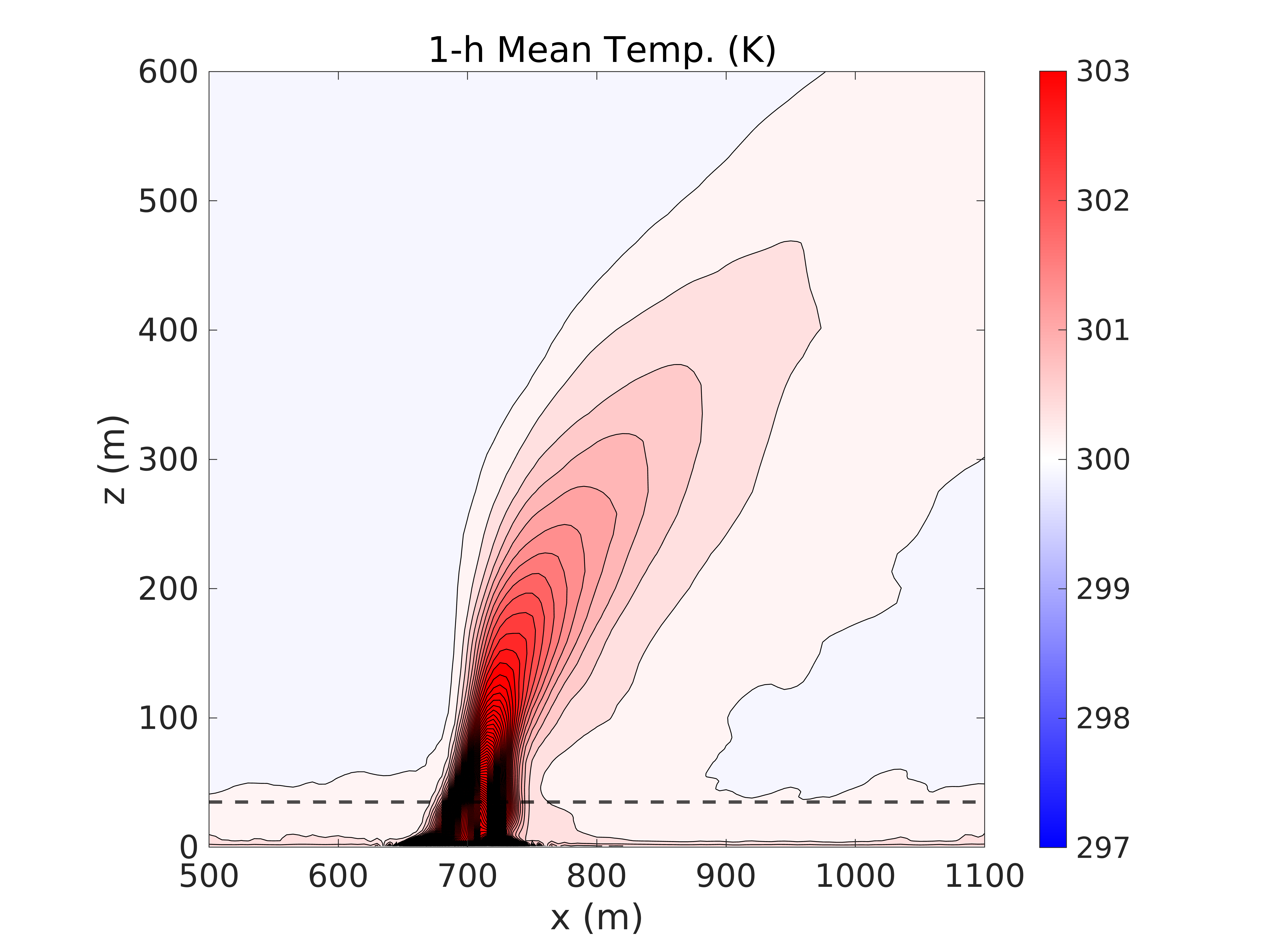}  
     & \includegraphics[scale=0.35]{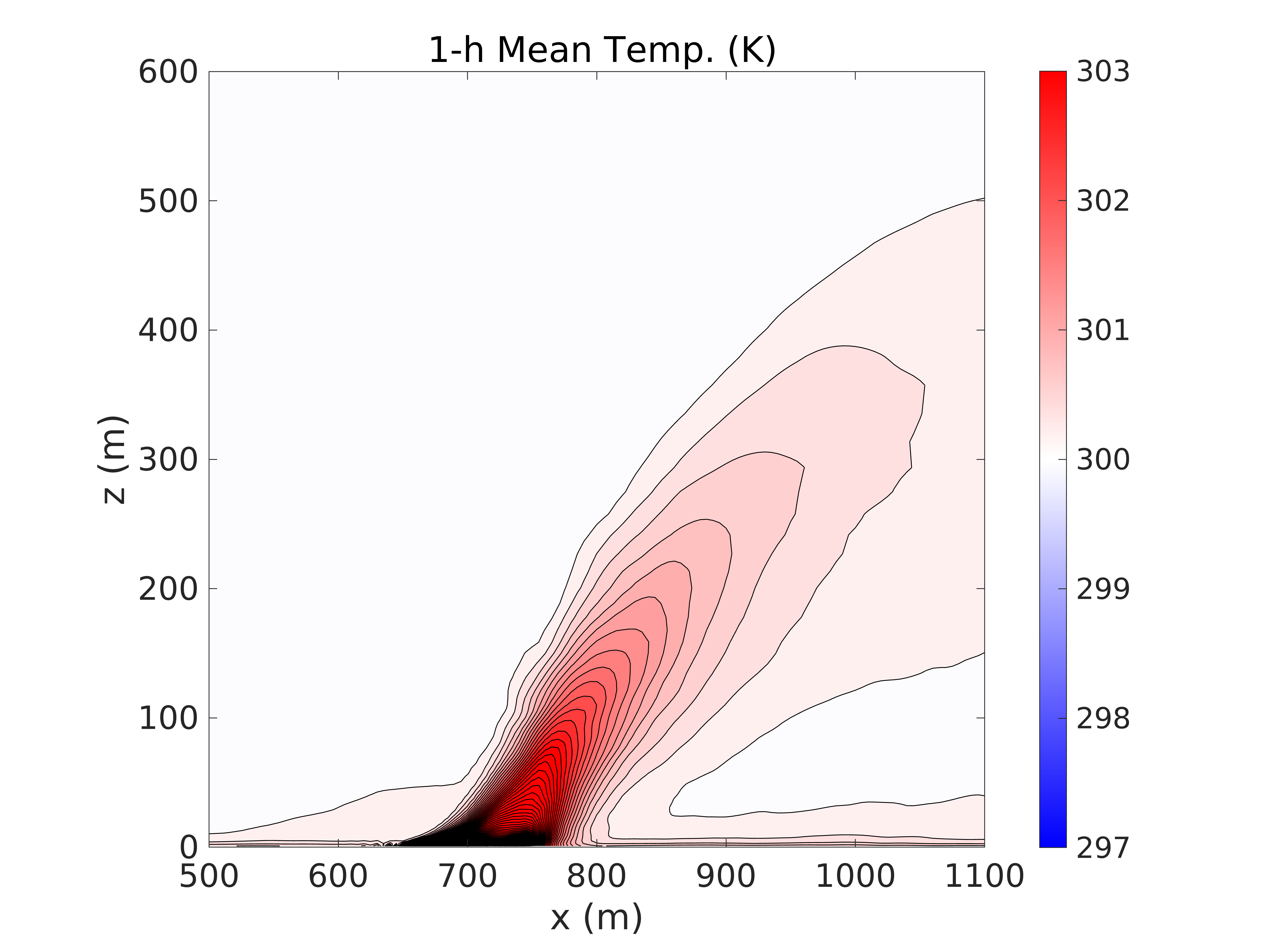} \\
         & 
    \end{tabular}
    \caption{Contours of the 1-h mean temperature plotted on the $XZ$ plane $y=500$\,m (passing through the center of the heated patch) for the (a) canopy and (b) no-canopy cases. Dashed horizontal line in (a) represents $z=h_\text{c}$}
    \label{fig_sublayer_temp}
\end{figure}

\begin{figure}[h!]
    \centering
\begin{tabular}{cc}
    \includegraphics[scale=0.41]{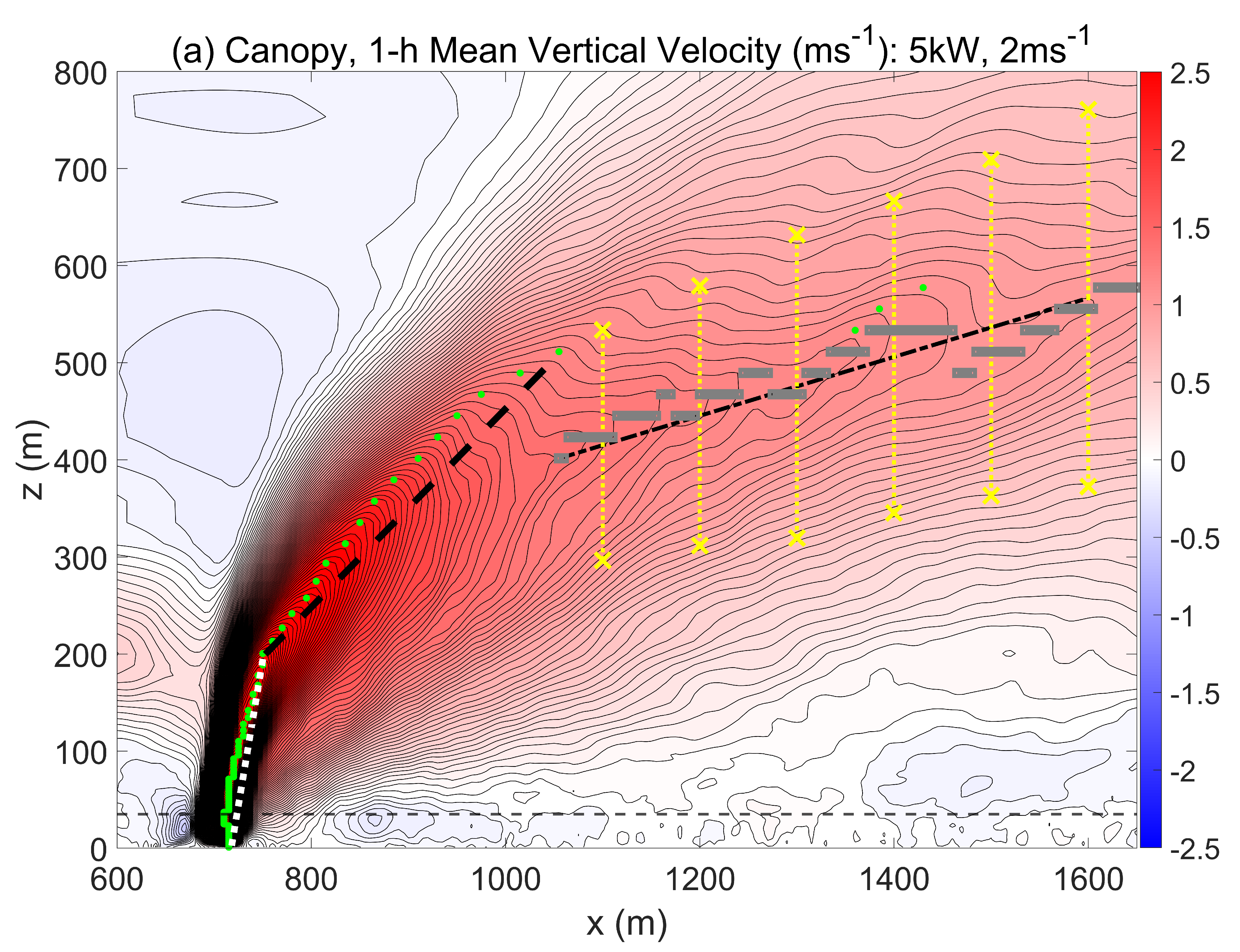}& 
     \includegraphics[scale=0.41]{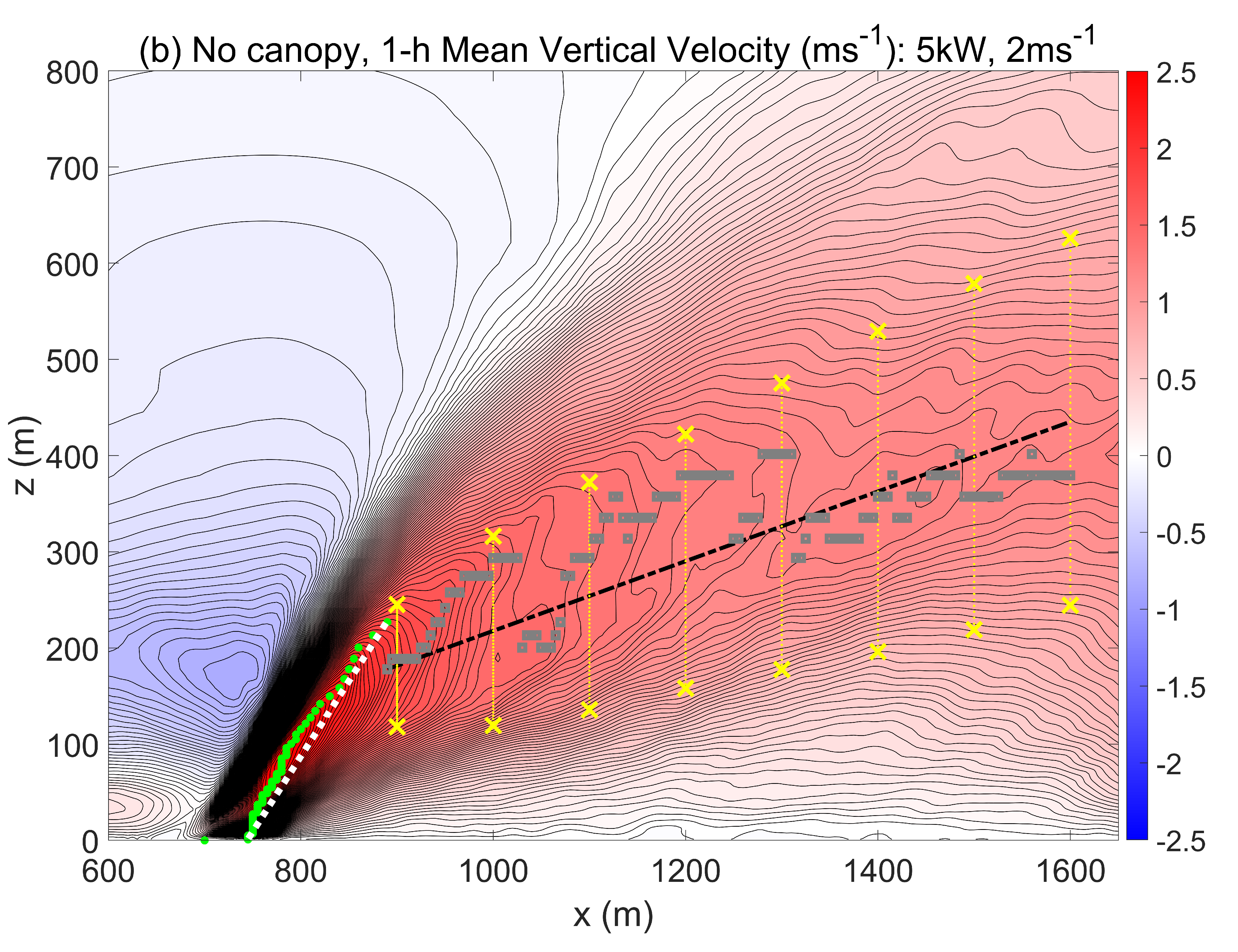}
\end{tabular}
    \caption{Color contours of the 1-h mean vertical velocity ($\overline{w}$) for the (a) canopy and (b) no-canopy cases. Grey points represent 1-h mean plume centerline ($\mu$) in the far-field (Zone 2); vertical yellow dotted lines represent $\mu\pm s_t$ at multiple locations in the far-field (Zone 2); green dots represent the plume centerline in Zones 1 and 1(B) (in the canopy case), while white slanted dotted lines represent the plume centerline inclination. In (a), the black dashed line represents the inclination of the plume centerline in the transition zone (Zone 1(B)). Trend-lines (dash-dotted) in Zone 2 are obtained using MATLAB's \textit{fminsearch}() for a linear fit to the oscillating mean plume centerline}
    \label{fig_wav_stds}
\end{figure}

Figure~\ref{fig_sublayer_temp} depicts the 1-h mean potential temperature ($\overline{\theta}$) on the streamwise ($XZ$) plane passing through the center of the square surface patch ($y=500\,$m), for the canopy and no-canopy cases. It is observed from the temperature contours that the warm core of the plume is more vertical in the canopy case while it deflects more strongly from the vertical in the no-canopy case. Despite the hundred-fold difference in heat flux between the heated and the ambient surface, higher temperature relative to the ambient is only observed close to the plume core ($z<200$\,m) with temperature excursions from the ambient weakening with height, becoming negligible around $z=500$\,m. 
More information can be gathered from Fig.~\ref{fig_wav_stds}, which depicts the 1-h mean vertical velocity ($\overline{w}$) on the same $XZ$ plane 
for the canopy and no-canopy cases, along with the 1-h mean plume centerline. The mean plume centerline is tracked from the locus of the maximum 1-h mean vertical velocity ($\overline{w}_\text{max}$). In the region near the plume source (Zone 1), $\overline{w}_\text{max}$ is obtained from the maximum $\overline{w}$ across a horizontal transect in this plane at different heights. Further downstream from the plume source (Zone 2),  $\overline{w}_\text{max}$ is obtained from the maximum $\overline{w}$ across a vertical transect in this plane at different streamwise locations. The transition between the two zones is determined by tracking a sudden change in the local plume centerline slope beyond a certain threshold, from the green dots in Zone 1, as seen from Fig.~\ref{fig_wav_stds}. In the canopy case, the mean plume centerline slope in Zone 1 appears to be steeper due to the canopy drag on the plume. An additional zone above the canopy sublayer exists near the surface (Fig.~\ref{fig_wav_stds}(a)), where the mean plume centerline inclination transitions from the steeper inclination near the surface to a more gradual one before the onset of the far-field region. We refer to this zone as Zone 1(B). Note that the lines in Zones 1 and 1(B) depict the inclination in an approximate sense. The lines in Zone 2 are obtained using a simplex algorithm (MATLAB's \textit{fminsearch()} function) for a linear fit ($\mu$) to the oscillating mean plume centerline. 

The spatial standard deviation ($s_\mu$) of the far-field mean plume centerline altitude from the linear fit is approximately 20.40\,m in the canopy case and 49.95\,m in the no-canopy case. The 1-h mean plume centerline in the no-canopy case shows higher variability (more oscillatory behavior) compared to the canopy case. This seems to suggest that the far-field flow within the plume is more organized due to the canopy drag. 
Moreover, the transition to Zone 2 occurs at a higher altitude ($z\approx 511$\,m) in the canopy case as compared to the no-canopy case ($z\approx 227$\,m), which appears to be due to the aerodynamic effects of the canopy on the rising plume. Additionally, Fig.~\ref{fig_wav_stds} depicts the variation ($\mu\pm s_t$) in the linear fit to the 1-h mean plume centerline, obtained from the standard deviation of the time series of the plume centerline altitude ($s_t$) at multiple streamwise locations. These locations are 100\,m apart, i.e. $x=1100$\,m--1600\,m in the canopy case and $x=900$\,m--1600\,m in the no-canopy case. The time series of the plume centerline altitude at each streamwise location is obtained by tracking the height of the maximum instantaneous vertical velocity ($w_\text{max}$) at that streamwise location. The time variation in the plume centerline is observed to increase along the downstream direction in the far-field, although the variance at each streamwise location appears to be similar across the canopy and no-canopy cases. It can be noted that $s_t\sim\mathcal{O}$(100\,m) at $x=1100$\,m in the canopy case, suggesting high variability in the time-varying trajectory of the plume centerline owing to the highly turbulent interactions between the plume and the cross-wind. The visuals above suggest that although changes in the mean air temperature are spatially limited, for the current configuration, the variability in the flow field is more extensive, warranting further investigation in the following sections.

\subsection{Streamtubes and vortex tubes}\label{sect_stream_vort}
\begin{figure}[h!]
    \centering
    (a) Streamtubes (leeward-side POV)
    \includegraphics[scale=0.225]{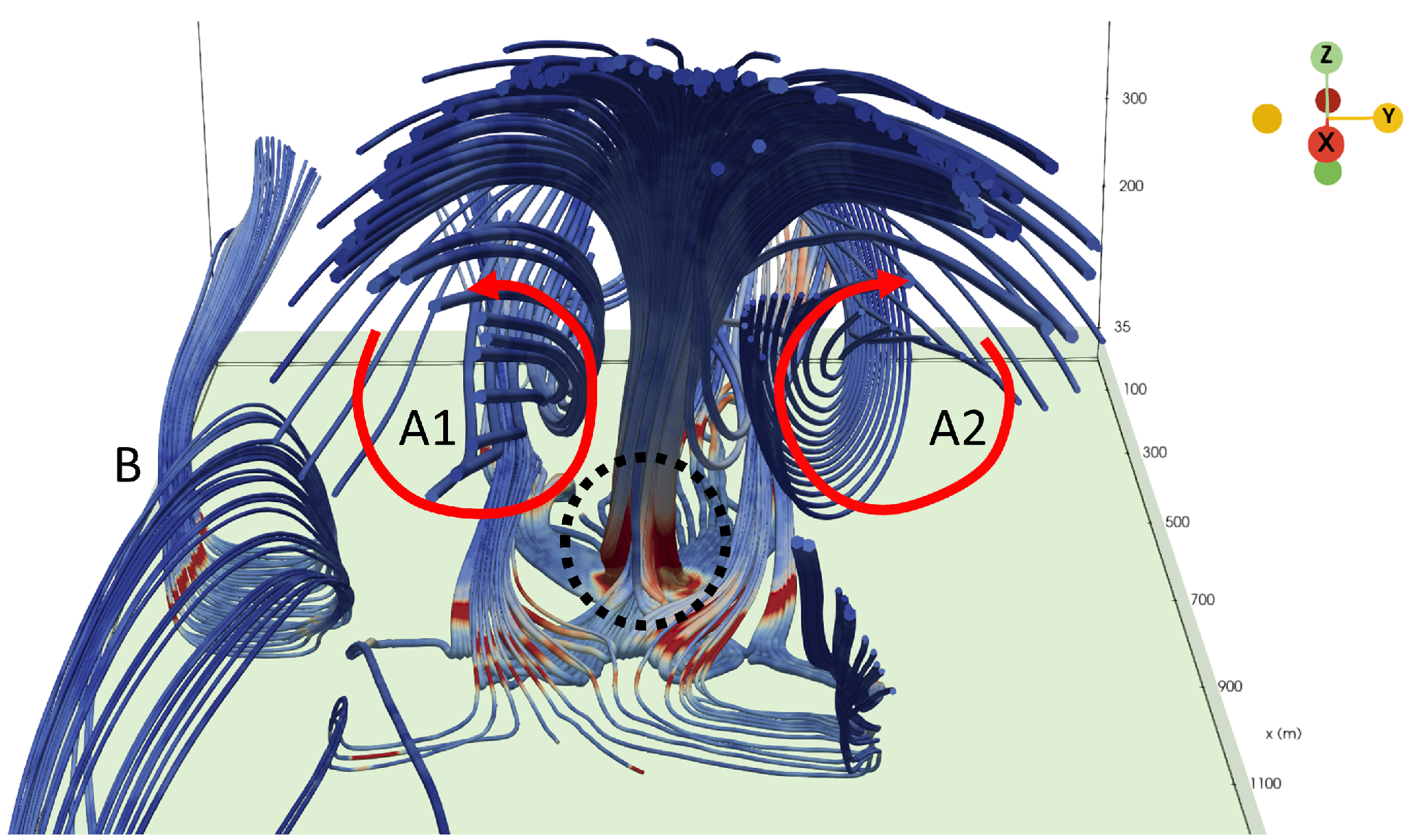}\\
    \vspace{15pt}
    \begin{tabular}{cc}
   (b) Streamtubes (near source)     &  (c) Streamtubes (tilted side view) \\
 \includegraphics[scale=0.2]{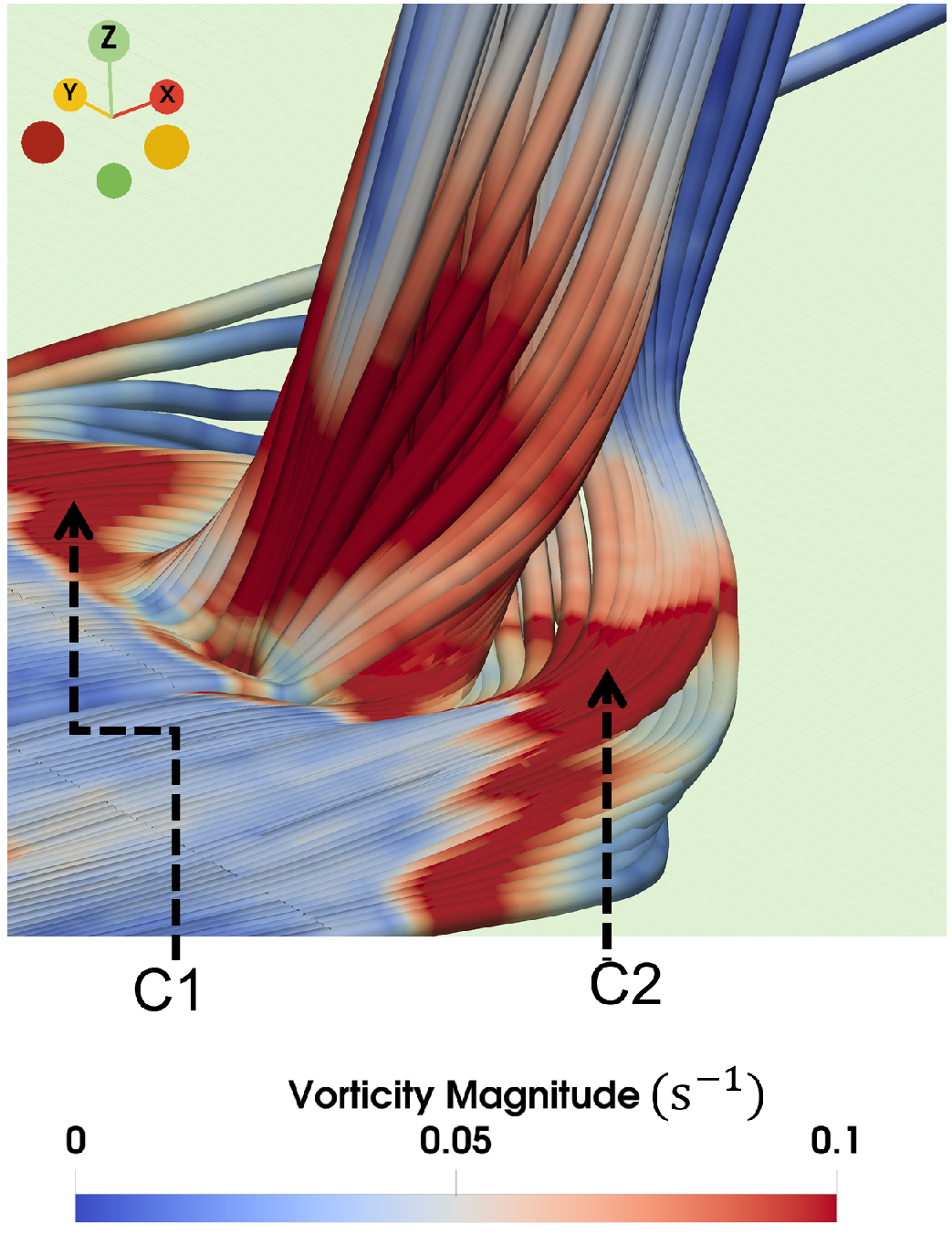}        & \includegraphics[scale=0.15]{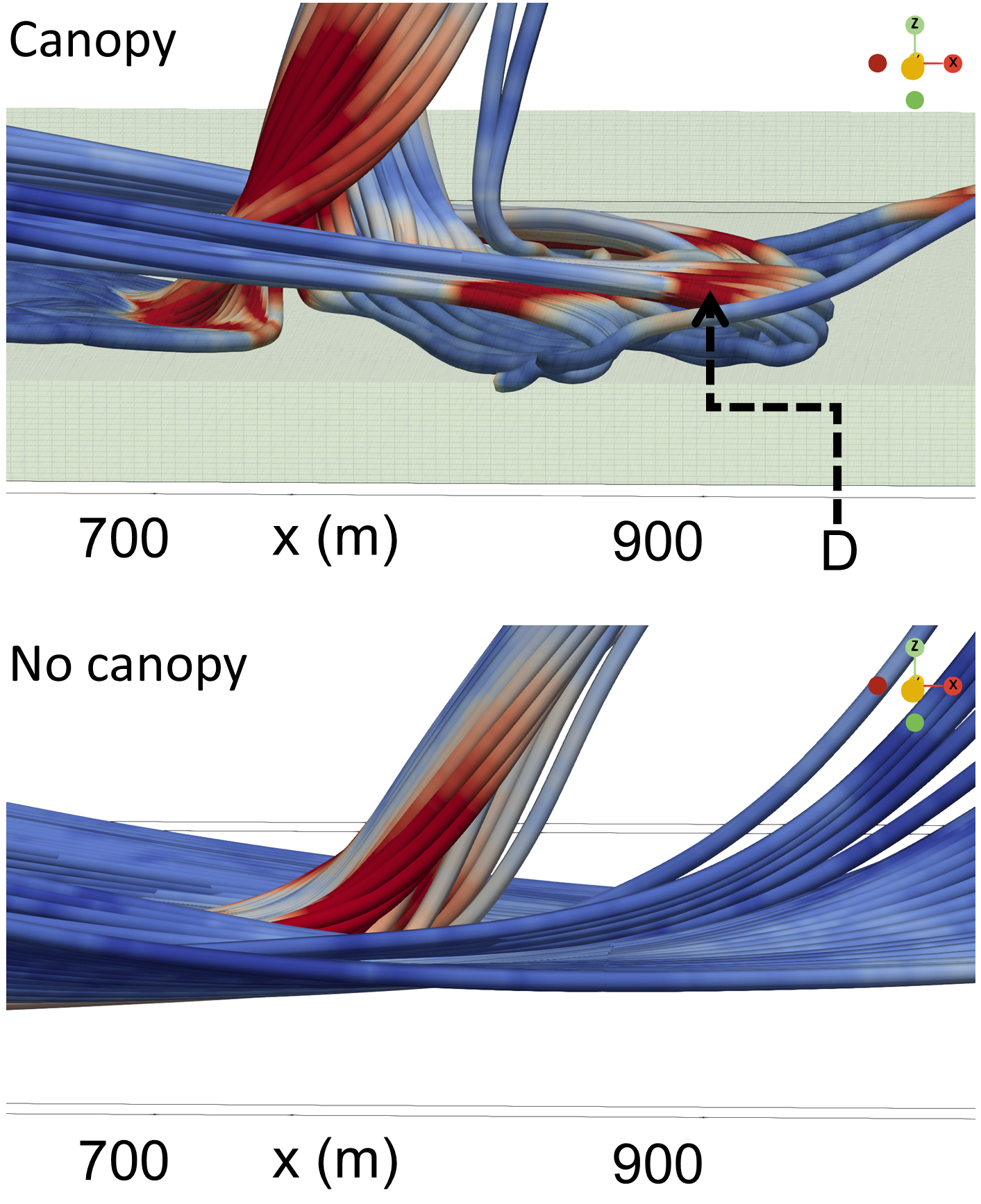}
    \end{tabular}
    \caption{3-D Streamtubes colored by vorticity magnitude ($\text{s}^{-1}$) (a) as observed from the downstream face of the domain and (b) demonstrating the presence of CVPs (C1, C2) near the source, in the canopy case. (c) A comparison of the structure of 3-D streamtubes between the canopy and no-canopy cases. with recirculating streamtubes. A1, A2--helical patterns along the plume edges; B-- secondary helical pattern; (D) recirculating streamtubes}
    \label{fig_streamtubes}
\end{figure}

Streamtubes are bundles of streamlines that are tangent to the local flow, and can give us insights into the spatial evolution of complex flow patterns across the flow field. Figure~\ref{fig_streamtubes}(a) depicts the streamtubes as viewed from the downstream side of the domain, colored by vorticity magnitude. Helical patterns are observed in streamtubes along the plume edges as the plume propagates downstream (A1, A2), representative of CVPs in the far-field region. 
On the lateral side of the plume, streamtubes encounter a \say{trough-like} downdraft region, followed by a region of updraft, exhibiting circulatory motion as they propagate downstream (B). These represent additional helical patterns induced outside the path of the plume, in the flow field, by the plume presence. Close to the source of buoyancy (enclosed in a dashed line), a region of high vorticity is observed. We zoom in on this region and view it from an alternative angle in Fig.~\ref{fig_streamtubes}(b). The high vorticity magnitude is found to correspond to streamtubes near the source, on either side, which rotate inward toward the source on the leeward side, resulting in CVPs (C1, C2), with their rotational axes aligned predominantly in the vertical direction ($\pm z$).

\begin{figure}[h!]
    \centering
    (a) Streamtubes: canopy (left) and no-canopy (right)\\
    \includegraphics[scale=0.22]{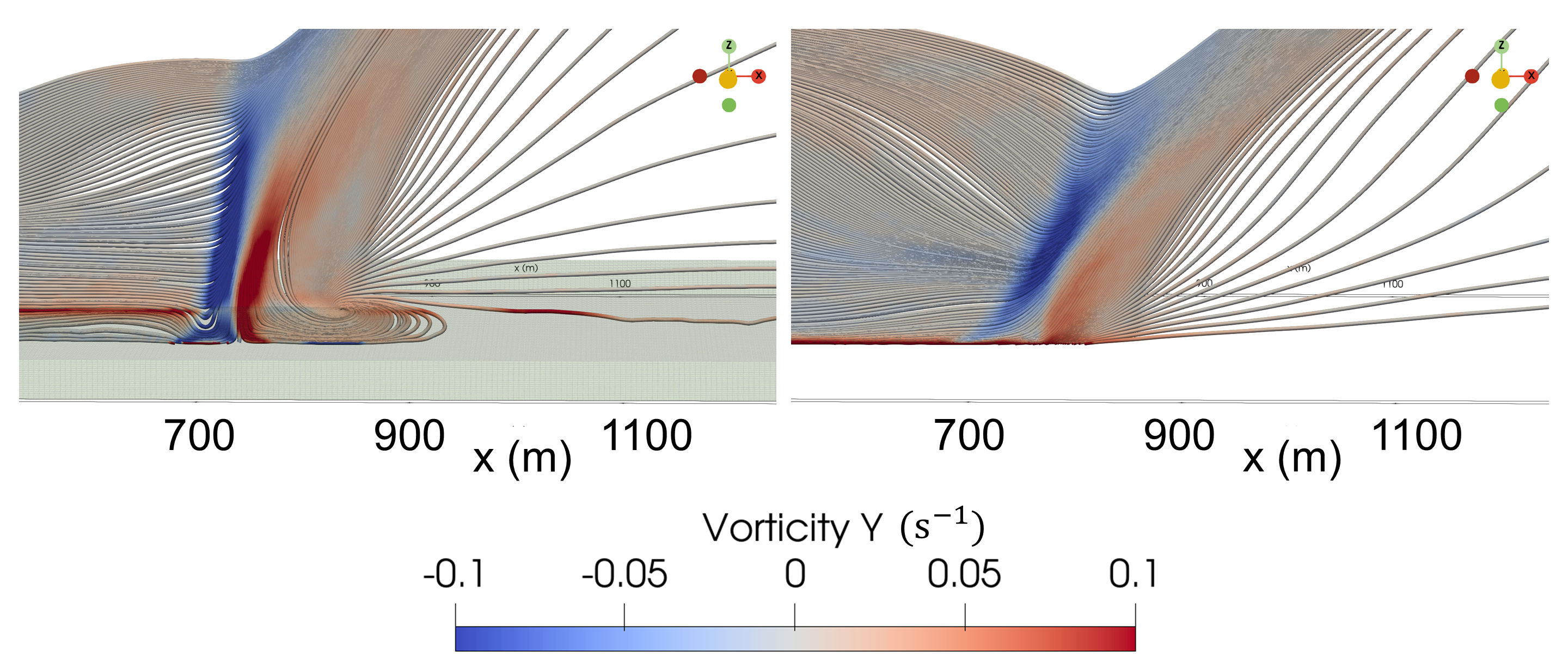}       \\
    \vspace{5pt}
    (b) Q-criterion: canopy (left) and no-canopy (right)
    \includegraphics[scale=0.22]{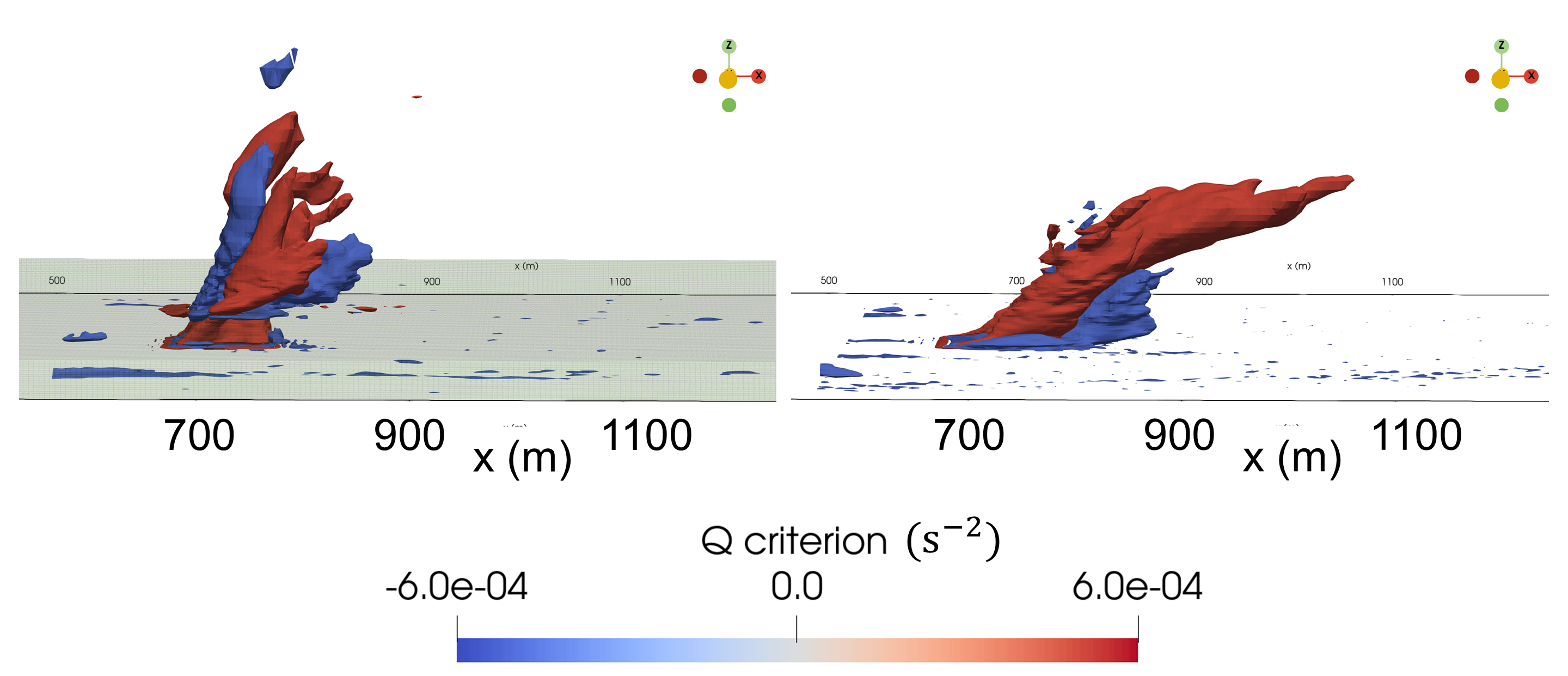} 
    \caption{(a) Streamtubes in the $XZ$ plane passing through the center of the surface patch and (b) isosurfaces of the Q-criterion shown for the canopy and no-canopy cases}
    \label{fig_qcrit}
\end{figure}

A comparison of streamtubes on the leeward side of the source between the canopy and no-canopy cases is made in Fig.~\ref{fig_streamtubes}(c). Two main differences can be observed. Recirculating streamtubes (D) are observed near the canopy top on the leeward side. These streamtubes represent flow that penetrates the canopy as they recirculate and the subsequent entrainment into the buoyant plume from the downstream side. Such recirculating streamtubes are absent in the no-canopy case. 
We explore this further from the streamtubes in the $XZ$ plane passing through the center of the surface patch, shown for the canopy and no-canopy cases, in Fig.~\ref{fig_qcrit}(a). In the canopy case, streamlines suggest flow that penetrates the canopy from both the upstream and downstream sides before being entrained by the plume updrafts. A recirculation zone is created on the leeward side, which results in a strong adverse pressure gradient as the pressure increases upstream of the zone. As the upstream flow approaches the plume, there is a decrease in horizontal momentum followed by entrainment into the plume core. Owing to this decrease, the plume tilt is relatively weaker in the presence of the canopy. In contrast, entrainment into the plume from the upstream side appears to be stronger, with high horizontal momentum, in the no-canopy case. Moreover, streamtubes from aloft are seen to impinge upon the plume on the upstream side. These dynamics result in a steeper tilt in the plume in the absence of a canopy at the surface. 

The streamtubes are colored by the cross-stream ($y$) component of vorticity ($\omega_y$) in Fig.~\ref{fig_qcrit}(a). In both cases, $\omega_y < 0$ on the windward side and $\omega_y > 0$ on the leeward side of the plume. It can be observed that the magnitude of $\omega_y$ increases within the plume, though it appears to be higher in the canopy case on both the windward and leeward sides. In the no-canopy case, the magnitude of $\omega_y$ is higher on the upstream side. The regions of increased $|\omega_y|$ in both cases are associated with stronger deflection of the streamlines and increased shear. 
To differentiate between regions of high vorticity and high strain, we plot iso-surfaces of the Q-criterion in Fig.~\ref{fig_qcrit}(b) for both canopy and no-canopy cases. The Q-criterion is given by 
\begin{equation}
\begin{split}
    Q = \frac{1}{2}(||\Omega||^2 - ||S||^2),~\text{where}~    \Omega = \frac{1}{2}\left(\frac{\partial u_i}{\partial x_j} - \frac{\partial u_j}{\partial x_i}\right)~\text{and}~S = \frac{1}{2}\left(\frac{\partial u_i}{\partial x_j} + \frac{\partial u_j}{\partial x_i}\right).
\end{split}
\end{equation}
Here, $\Omega$ is the rotation rate tensor and $S$ represents the strain rate tensor. In Fig.~\ref{fig_qcrit}(b), the dark, solid red regions ($Q>0$) indicate rotation-dominated regions, whereas dark, solid blue regions ($Q<0$) indicate strain-dominated regions. Note that, while the point of view is the same as in Fig.~\ref{fig_qcrit}(a), the isosurfaces are 3-D. The rotation-dominated regions correspond to regions of twisting and rotating streamtubes on either side of the plume near its source in both canopy and no-canopy cases. In the no-canopy case, the rotation-dominated isosurfaces are more tilted and stretched in the downstream direction, representing the tilting and stretching of vorticity by the entrained wind. Strain-dominated regions are observed on the windward and leeward sides of the plume in the canopy case, suggesting the presence of strong shear in these regions. On the windward side, the increased strain rate appears to be associated with the deformation of fluid elements as the cross-flow impinges upon the plume and adjusts to its strong, buoyant updrafts. Conversely, on the leeward side, the increased strain rate is associated with the deformation of fluid elements as the flow transitions from the influence of strong updrafts to that of the cross-wind. Increased strain is more prominent on the leeward side in the no-canopy case.\\

\begin{figure}[h!]
    \centering
    \includegraphics[scale=0.225]{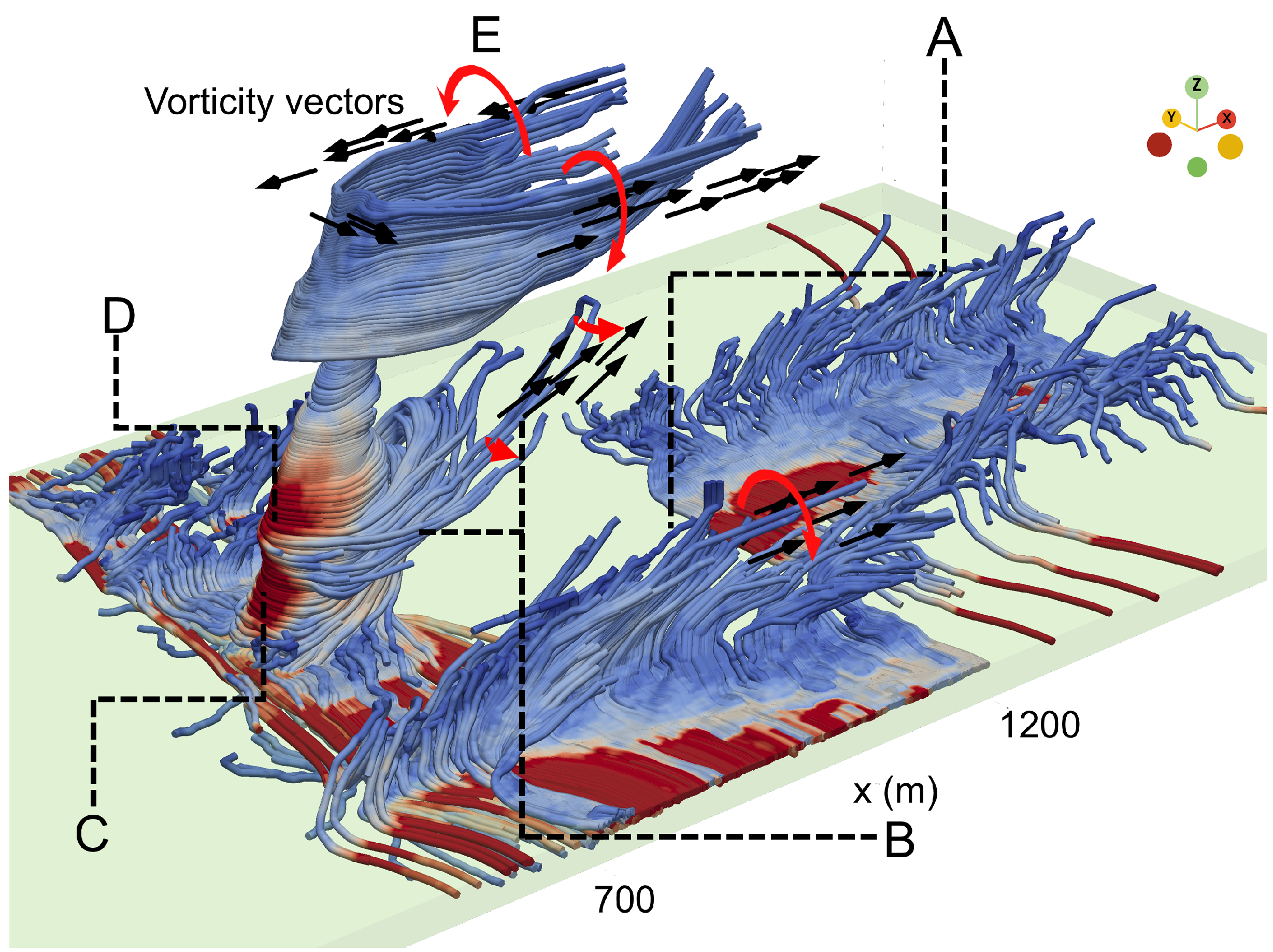} \\
    \includegraphics[scale=0.12]{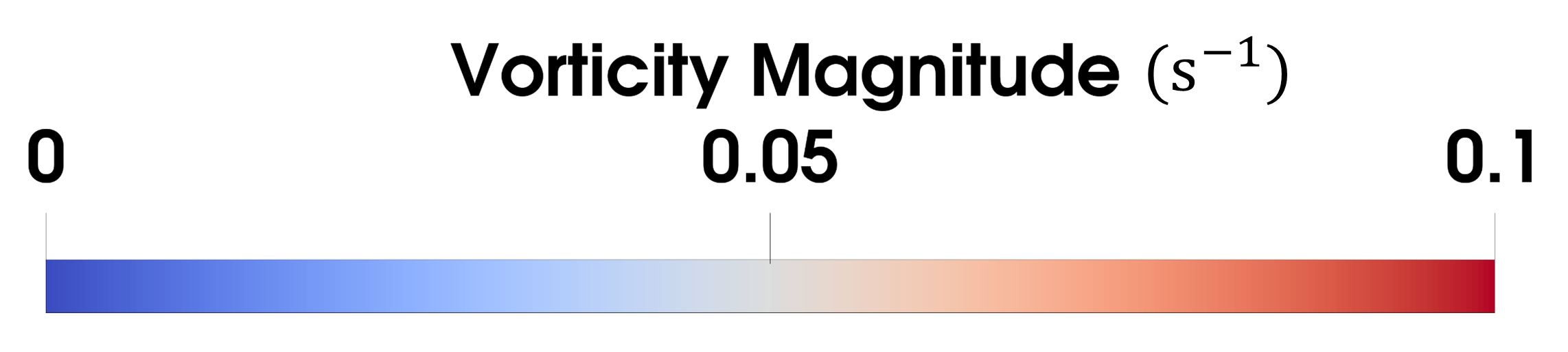}
    \caption{Vortex tubes colored by vorticity magnitude ($\text{s}^{-1}$). A--Vortex tubes influenced by updrafts, cross-wind; B--Plume-modulated and tilted hairpin vortices; C--Spiraling vortex tubes in plume; D--Vorticity intensification 
at plume windward edge; E-- CVPs propagating
downstream. Black solid arrows represent vorticity vectors and red arrows the corresponding circulation structures. }
    \label{fig_vort_tubes}
\end{figure}

A comprehensive picture of the circulation patterns in the flow field can be obtained from vortex tubes, i.e. bundles of lines tangent to the local vorticity vectors. Figure~\ref{fig_vort_tubes} represents vortex tubes colored by vorticity magnitude ($\text{s}^{-1}$), both within and outside the plume, in the presence of the canopy. It is observed that vortex tubes, near the canopy top, aligned in the direction transverse to the domain are lifted by plume-induced secondary updrafts (A) in the region transverse to the plume. These vortex tubes are then aligned in the streamwise direction by the cross-wind; the associated vorticity vectors, obtained from a right-hand thumb rule, suggest circulations that direct the flow along a helical pattern, as shown by the streamlines above (B in Fig.~\ref{fig_streamtubes}(a)). On the leeward side of the plume, vortex tubes are tilted and stretched (B), via interaction with buoyant updrafts and the cross-wind; the corresponding vorticity vectors suggest hairpin-like vortical structures. Within the plume, vortex tubes are seen to spiral inwards into the plume core (C), resembling ring-like vortical structures and demonstrating entrainment of the surrounding vortical motions into the plume core. This is similar to vorticity entrainment as seen in tornado formation \cite{rotunno2024recent}, during which entrained circulation patterns respond to updrafts, resulting in whirl-like phenomena; however, the intensities of such whirl-like circulations in the current case may differ from those associated with tornado formation. On the windward side, these ring-like vortex tubes are pressed toward each other in regions where the vorticity magnitude is relatively higher (C, D). This increase in vorticity magnitude at regions where vortex tubes are more concentrated has also been observed along the fire-front in an experimental study on small-scale surface fires \cite{desai2021investigating} spreading from a point ignition in calm ambient wind conditions. Finally, in the region where the plume interacts with the crosswind aloft, the circulation structures corresponding to the vortex tubes and associated vorticity vectors suggest the presence of CVPs that propagate downstream (E), which can be corroborated by the helical streamtubes seen in Fig.~\ref{fig_streamtubes}(a).

We take the opportunity at this point to comment on the extension of the vortical patterns observed from the current setup to intense fires. Church et al. \cite{church1980intense} used Meteotron, an array of 105 fuel oil burners in a 140\,m$\times$140\,m burn area, with a total heat output of approximately 1000\,MW to generate a hot smoke plume from an intense experimental fire, which interacted with surface winds ranging from 3--5\,ms\textsuperscript{-1}. Although the heat output in the current case is 20 times less intense compared to their case, the vortical patterns observed were similar, including the presence of CVPs near the source, horizontal vortex tubes morphing into hair-pin vortices on the leeward side of the plume, and vortex rings within the rising plume. Moreover, burners produced flames approximately 5--10\,m in length, while flames are absent in the current case, indicating the relevance of our low-complexity, no-flame setup in the exploration of fire-induced turbulent flow features. 

\begin{figure}
    \centering
    \includegraphics[scale=0.225]{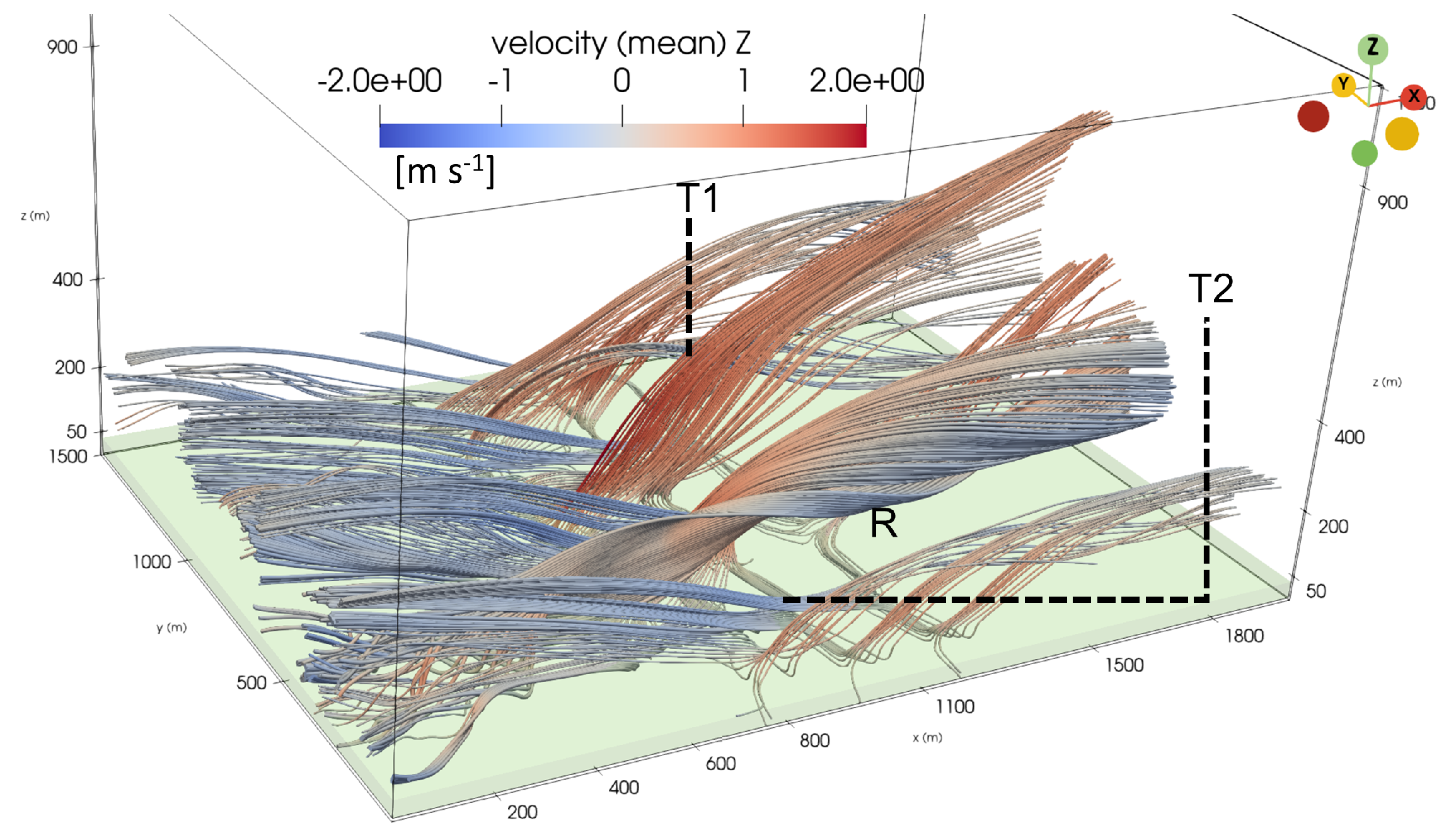}
    \caption{Streamtubes for a high-aspect ratio heat source colored by $\overline{w}$. T1--tower-like regions; T2--trough-like regions; R--rotational/helical patterns}
    \label{fig_fireline_stream}
\end{figure}

Next, we investigate the effect of the heat-source geometry. Figure~\ref{fig_fireline_stream} depicts streamtubes describing the interaction of the same cross-wind conditions as above with a buoyant plume emanating from a high-aspect ratio ($L_\text{y}/L_\text{x} = 25$) heat source resembling a line fire at the surface beneath a canopy. The size of the heat source is 20\,m$\times$500\,m, so that its area, and hence the total power output ($50$\,MW), is the same as that for the square-shaped heat source described above. Certain similarities exist with the structures seen from the square heat source, i.e. the presence of the secondary helical pattern on the transverse side of the source and helical/rotational patterns in the plume as it propagates downstream. In addition, alternating tower-trough-like structures (T1, T2) are observed in the plume. In the trough-like (T2) regions, the streamtubes resemble forward burst-like flow patterns that potentially transport heat downstream. In the tower-like regions (T1), the rotating streamtubes describe a helical pattern (R) in the plume as it propagates downstream. The circulation patterns in the plume are aligned in the transverse direction such that the troughs correspond to the downwash regions and the towers to the upwash regions. Such tower-trough-like regions and circulation patterns have been observed in laboratory experiments \cite{finney2015roleof} and FIRETEC simulations  \cite{banerjee2020identifying} on propagating flames as well. Again, the common features between our static heat source and propagating firelines show considerable promise for the current low-complexity setup to capture much of the plume dynamics observed from otherwise more resource-intensive simulations. 

\subsection{Momentum Fluxes and Quadrant Analysis}
\begin{figure}
    \centering
    \begin{tabular}{cc}
     \includegraphics[scale=0.2]{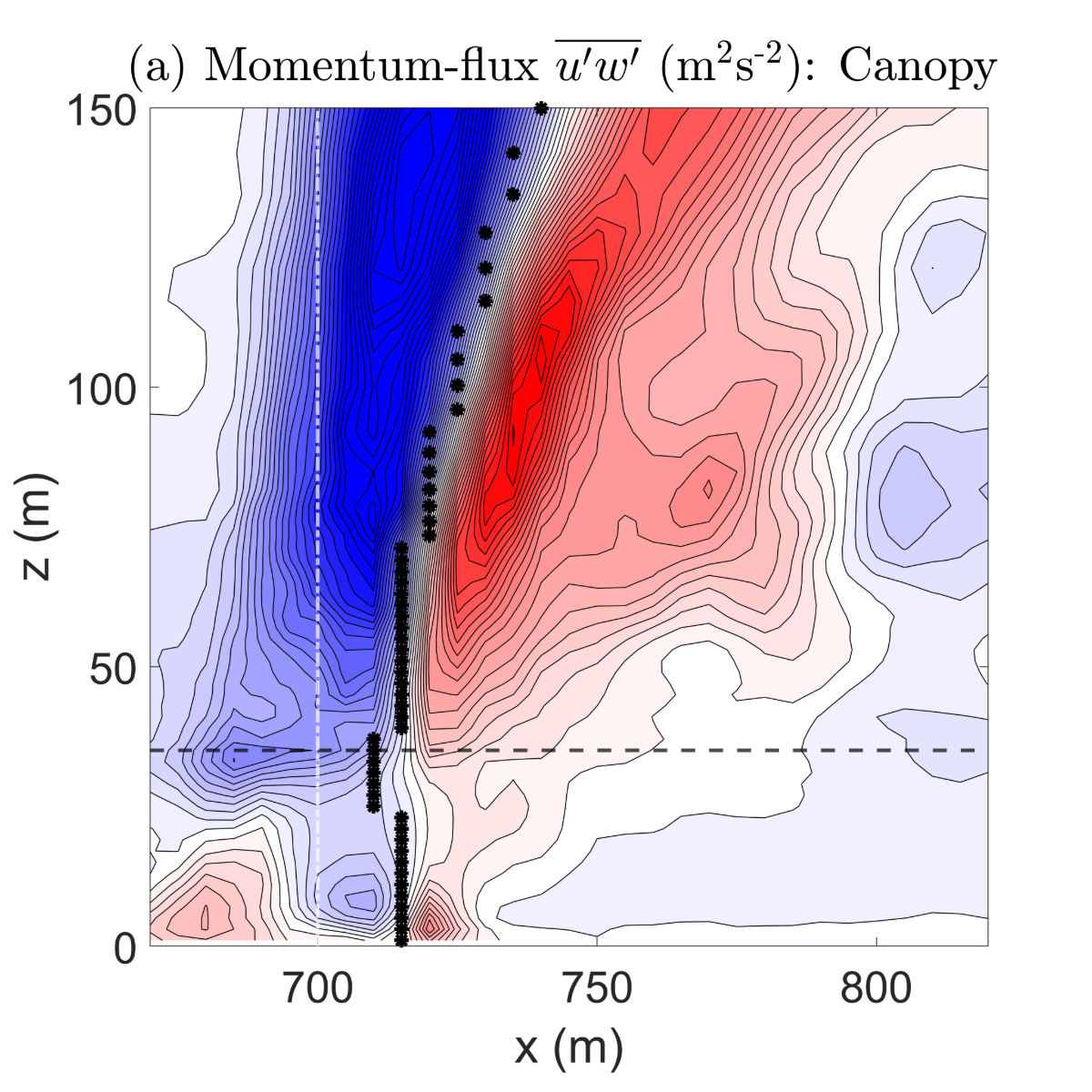}
        &  \includegraphics[scale=0.2]{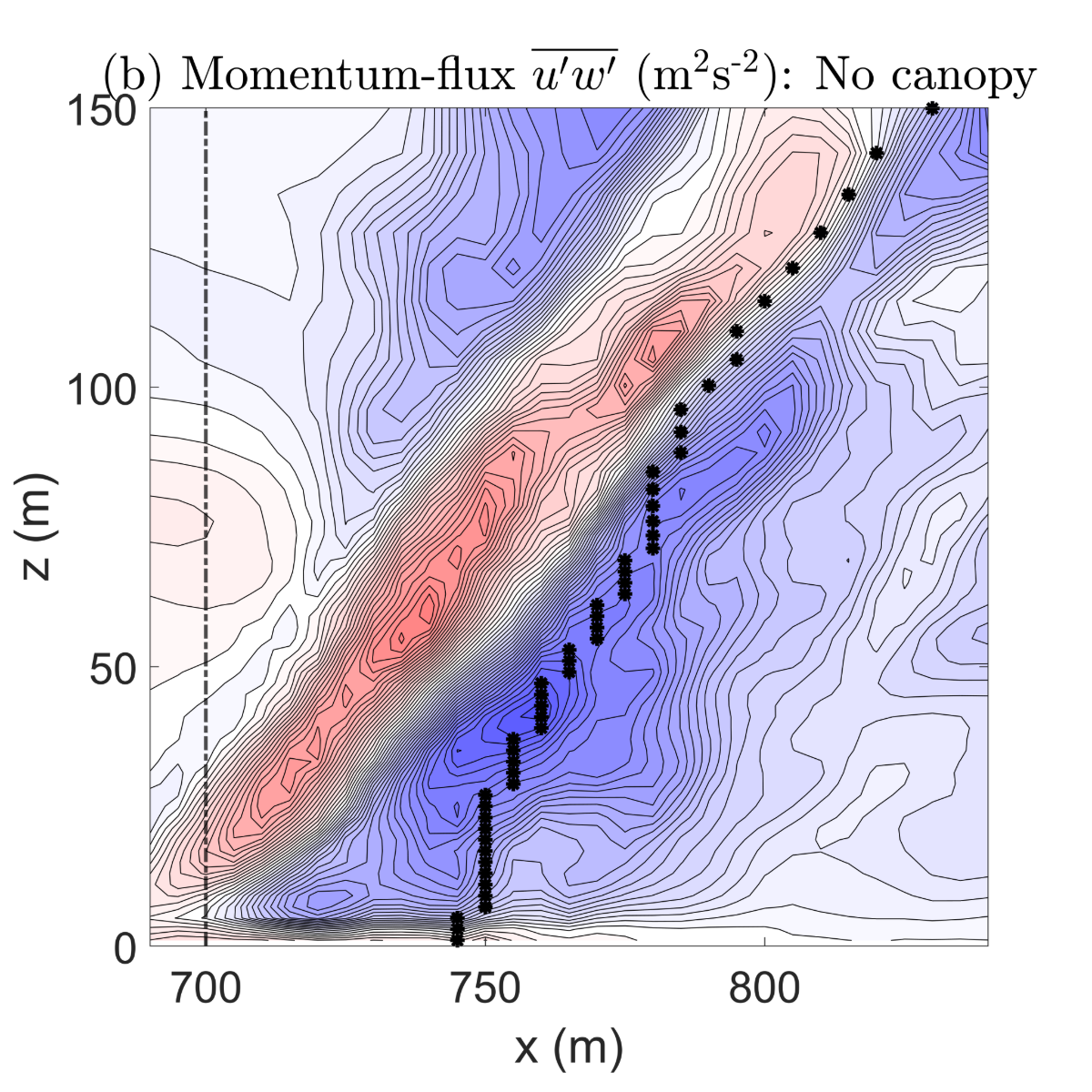}  \\
        \includegraphics[scale=0.055]{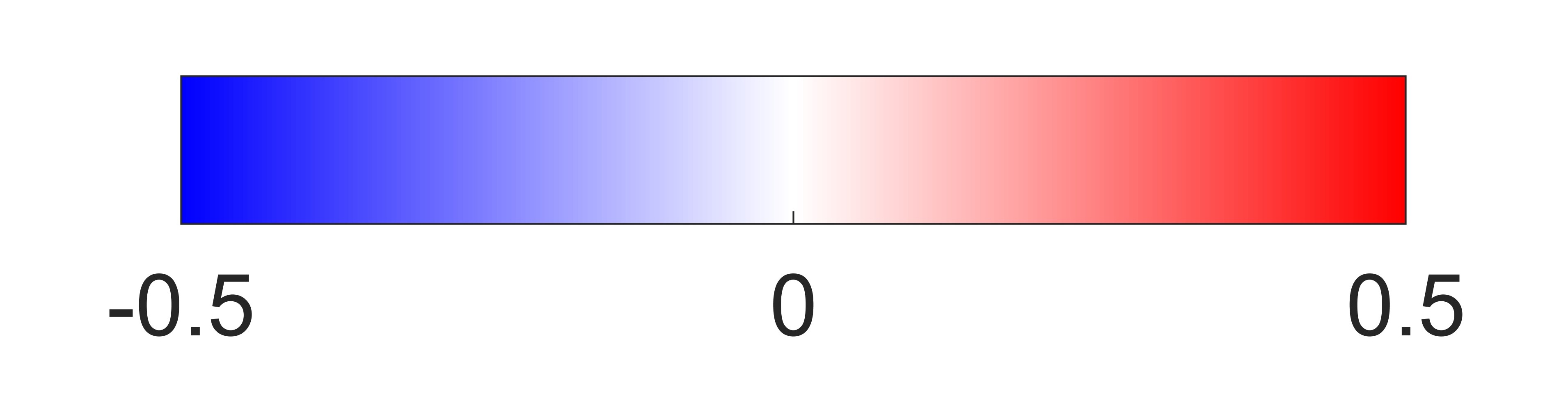} &  \includegraphics[scale=0.055]{net_uw_nocan_sublayer_colorbar.jpg} \\
        \includegraphics[scale=0.2]{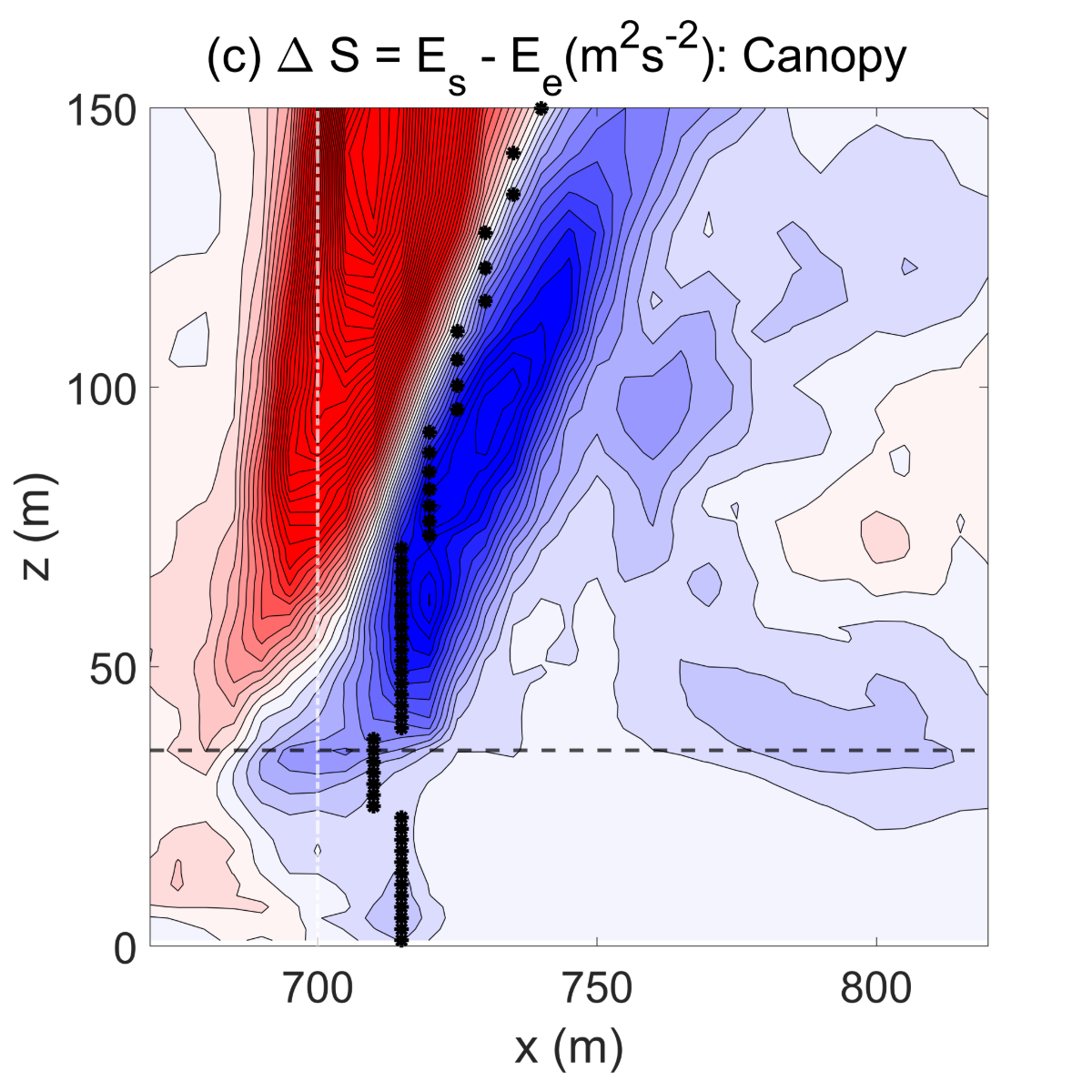} &  \includegraphics[scale=.2]{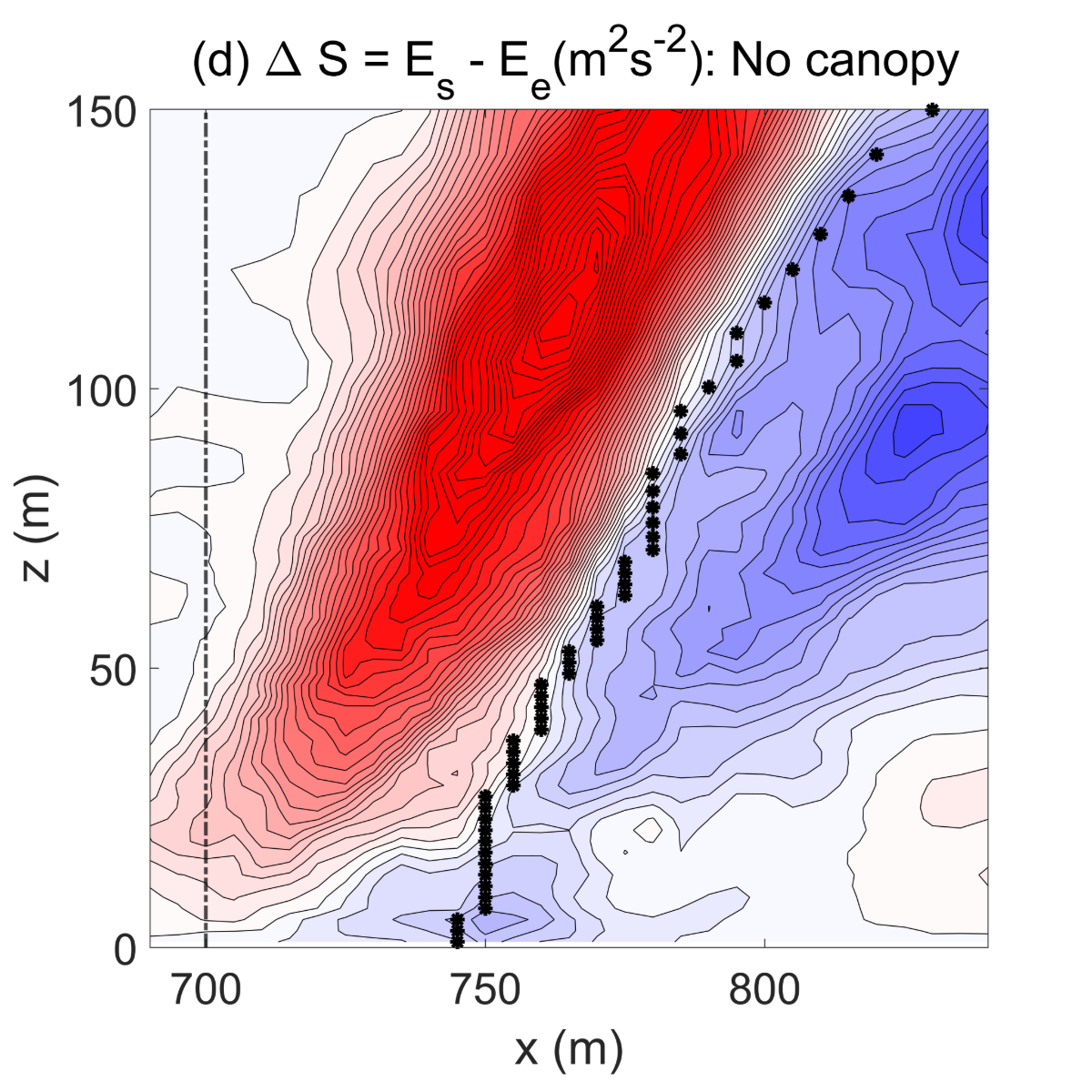}  \\
 \includegraphics[scale=.2]{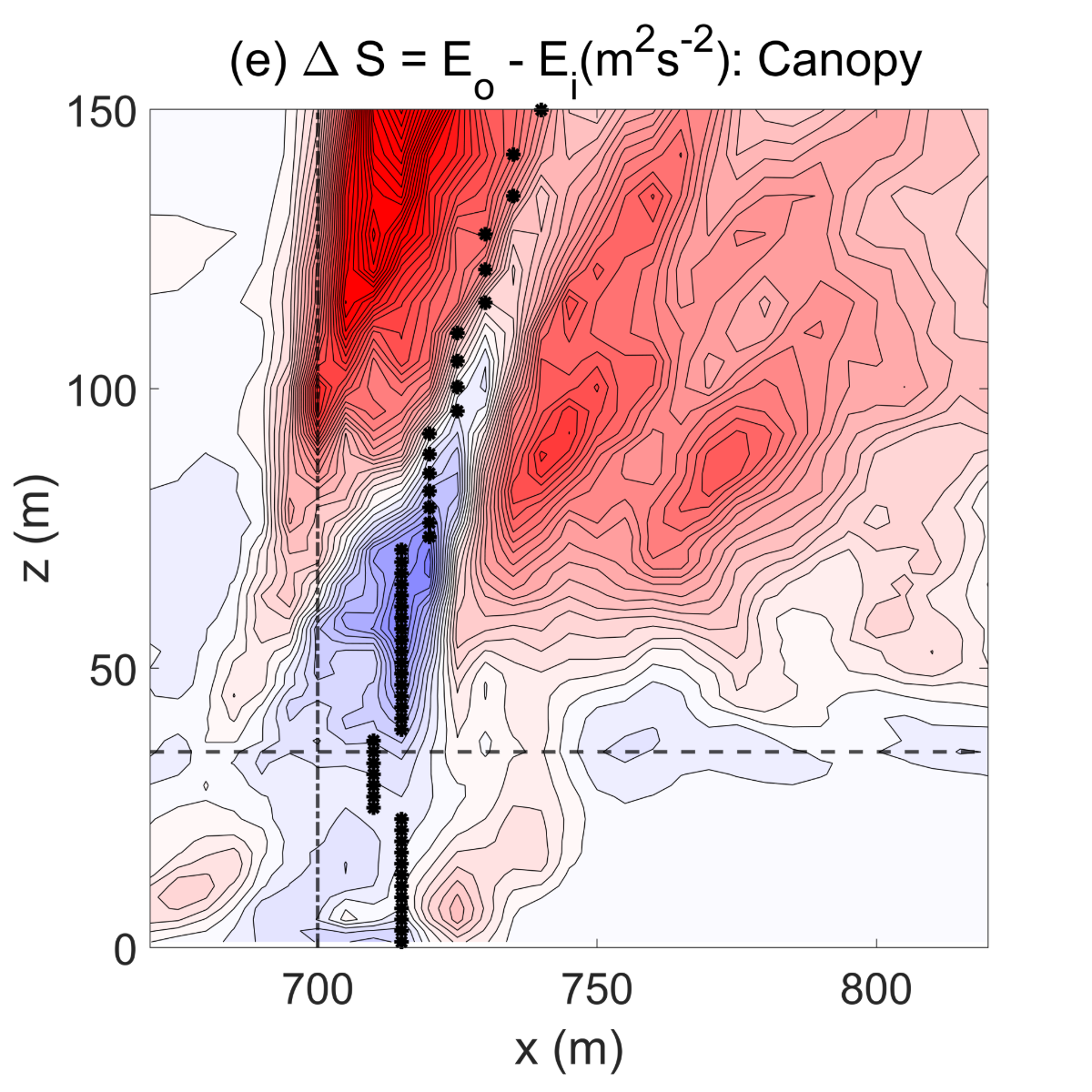} & \includegraphics[scale=.2]{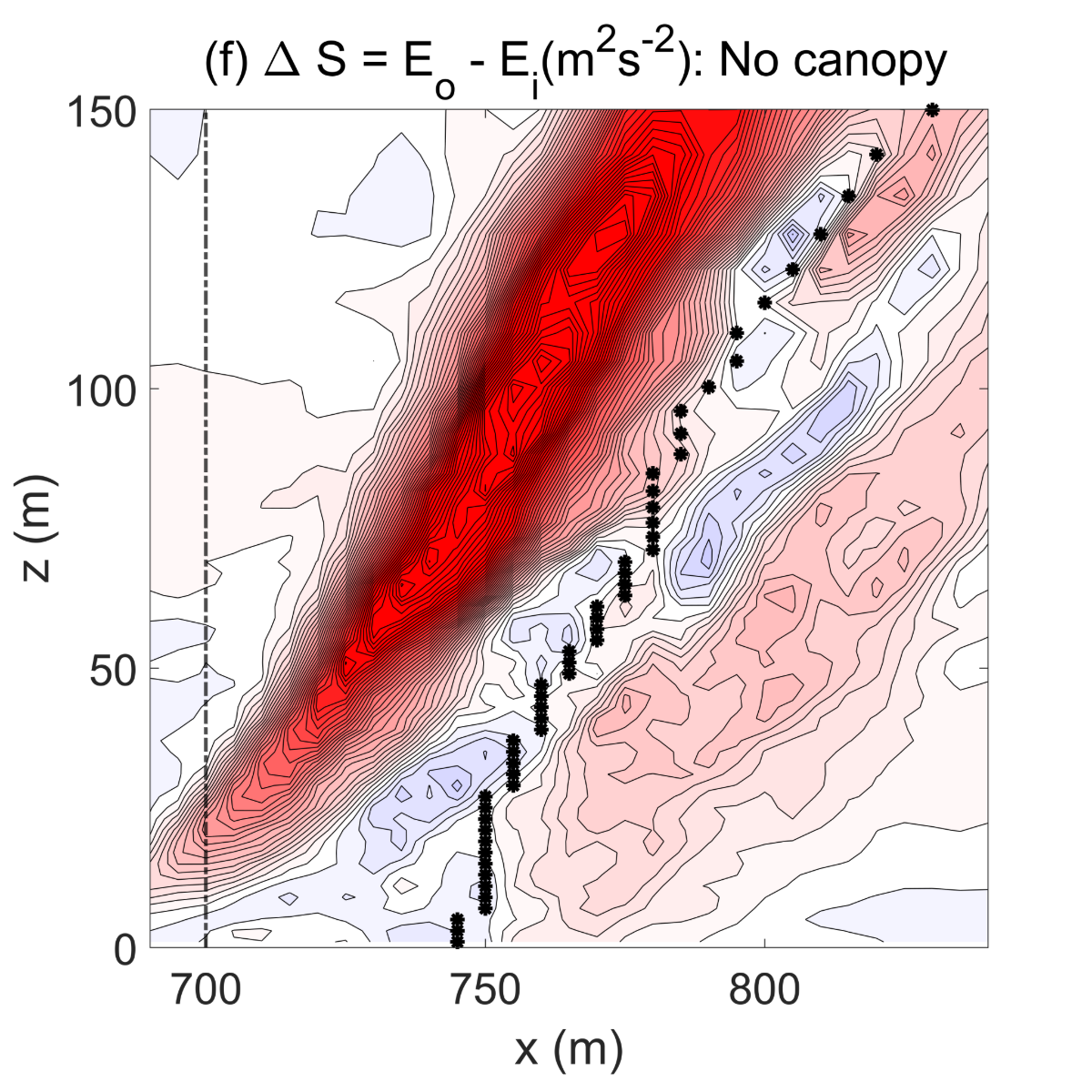}
    \end{tabular}
    \includegraphics[scale=.075]{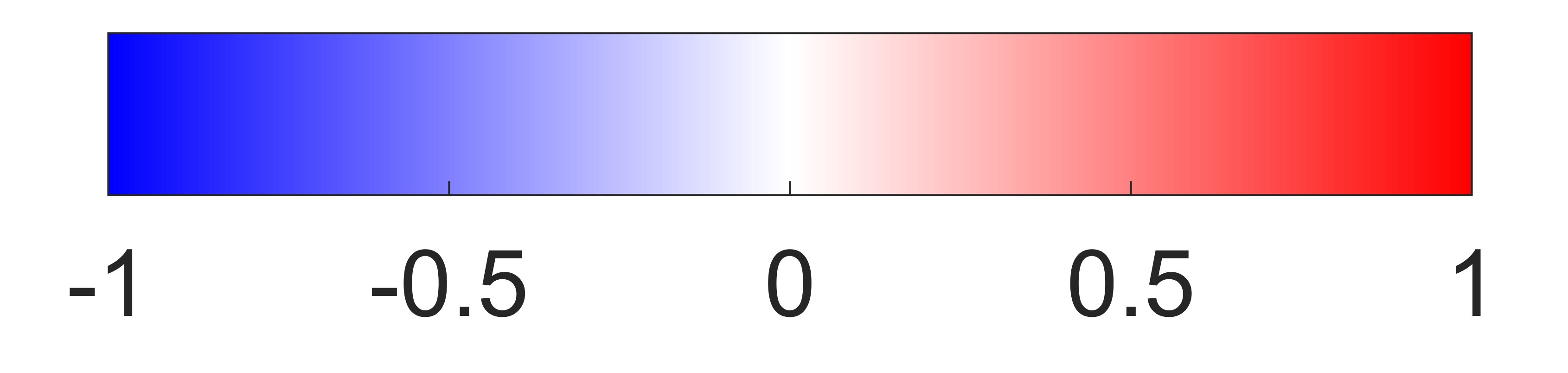}
    \caption{Turbulent momentum fluxes ($\overline{u'w'}$) in the (a) canopy and (b) no-canopy cases close to the surface. $\Delta S = E_\text{s}-E_\text{e}$ in the (c) canopy and (d) no-canopy cases. $\Delta S = E_\text{o}-E_\text{i}$ in the (e) canopy and (f) no-canopy cases. Black dots represent mean plume centerlines in Zone 1}
    \label{fig_uw}
\end{figure}

Thus far, we have discussed the features of the mean flow field. We now focus on insights that can be obtained from the turbulent momentum-flux events near the plume source and attempt to differentiate between the canopy and no-canopy cases in that regard. Figures~\ref{fig_uw}(a)--(b) depict the turbulent momentum flux, $\overline{u'w'}$, for the canopy and no-canopy cases, respectively, on the $XZ$ plane passing through the symmetry axis of the heat source, with the overbar representing the 1-h block mean as discussed in Sect.~\ref{sect_setup}. Mean plume centerlines from Zone 1 are also plotted for both cases. 
The regions where $\overline{u'w'}<0$ represent regions where momentum is transported to the surface, whereas regions where $\overline{u'w'}>0$ represent regions where momentum is injected into the atmosphere aloft. Momentum fluxes within the plume demonstrate spatial heterogeneity along with a noticeable structural difference between the canopy and no-canopy cases. In the canopy case, a clear distinction is observed between the nature of the momentum fluxes on the upstream and downstream sides of the mean plume centerline. Momentum is transported from the atmosphere aloft towards the plume source on the windward side. This region corresponds to the region where ambient wind loses its horizontal momentum as it approaches the plume, as noted in Sect~\ref{sect_stream_vort}. This region extends through the canopy, all the way to the surface. On the leeward side of the mean plume centerline, which remains relatively sheltered from the cross-wind, momentum is transported away from the source of buoyancy into the atmosphere aloft. 
In the no-canopy case, however, the plume region where momentum is injected into the atmosphere, by turbulent fluxes away from the buoyancy source is juxtaposed on either side by plume regions where momentum is transported to the surface/buoyancy source. Moreover, the intensity of the momentum fluxes appears to be higher in the canopy case.  

Through a quadrant analysis, momentum-flux events can be categorized into sweeps (downfluxes of high momentum, i.e. $u'>0, w'<0$), ejections (upfluxes of low momentum, i.e. $u'<0, w'>0$), outward interactions (upfluxes of high momentum, i.e. $u'>0, w'>0$), and inward interactions (downfluxes of low momentum, $u'<0, w'<0$). Of these, sweeps and ejections only contribute negative momentum fluxes, while outward and inward interactions only contribute positive momentum fluxes. Once categorized, the contribution to the net momentum flux from each of these quadrants ($E$) can be computed. Figures~\ref{fig_uw}(c)--(d) depict the difference between the contribution of sweeps and ejections ($\Delta S = E_\text{s}-E_\text{e}$) for the canopy and no-canopy cases, respectively. In both cases, ejections dominate sweeps on the windward side of the plume (red region), whereas sweeps dominate ejections on the leeward side (blue region). However, in the canopy case, the overall effect on the leeward side is dominated by positive momentum fluxes as seen from Fig.~\ref{fig_uw}(a), making the dominance of ejections on the windward side the more important observation. In addition, sweeps are observed to contribute more within the canopy volume on the windward side of the plume centerline.  
Additionally, Figs.~\ref{fig_uw}(e)--(f) depict the difference between the contribution of outward and inward interactions ($\Delta S = E_\text{o}-E_\text{i}$) for the canopy and no-canopy cases, respectively. 
A higher momentum-flux contribution from outward interactions compared to inward interactions is observed on the windward side of the plume centerline, particularly for the no-canopy case. In the canopy case, however, outward interactions also appear to dominate on the leeward side, which is a significant observation since the overall effect is dominated by positive momentum fluxes on the leeward side. 
Inward interactions are observed to prevail only over a thin region along the plume centerline in both the canopy and no-canopy cases. Within the canopy volume, inward interactions prevail on the windward side; however, these are overshadowed by negative momentum fluxes as seen in Fig.~\ref{fig_uw}(a); moreover, outward interactions prevail on the leeward side within the canopy volume (Fig.~\ref{fig_uw}(e)). Therefore, in the canopy presence, momentum appears to be transferred toward the canopy top primarily through upfluxes of low momentum (ejections), followed by downfluxes of high momentum (sweeps) toward the buoyancy source within the canopy, on the windward side of the plume centerline. Momentum is transferred from the buoyancy source to the atmosphere aloft on the leeward side through upfluxes of high momentum (outward interactions), both within and above the canopy subspace. The high contribution of sweeps and increased contribution from outward interactions below the canopy height, is also in agreement with observations from prescribed burns in forested environments \cite{heilman2021observations, heilman2021turbulent} as mentioned in Sect.~\ref{sect_intro}. In fact, the increased contribution from outward interactions often competes with the contributions from sweeps and ejections in both environments \cite{heilman2021observations}. 


\begin{figure}
    \centering
     \begin{tabular}{cc}
    \includegraphics[scale=0.17]{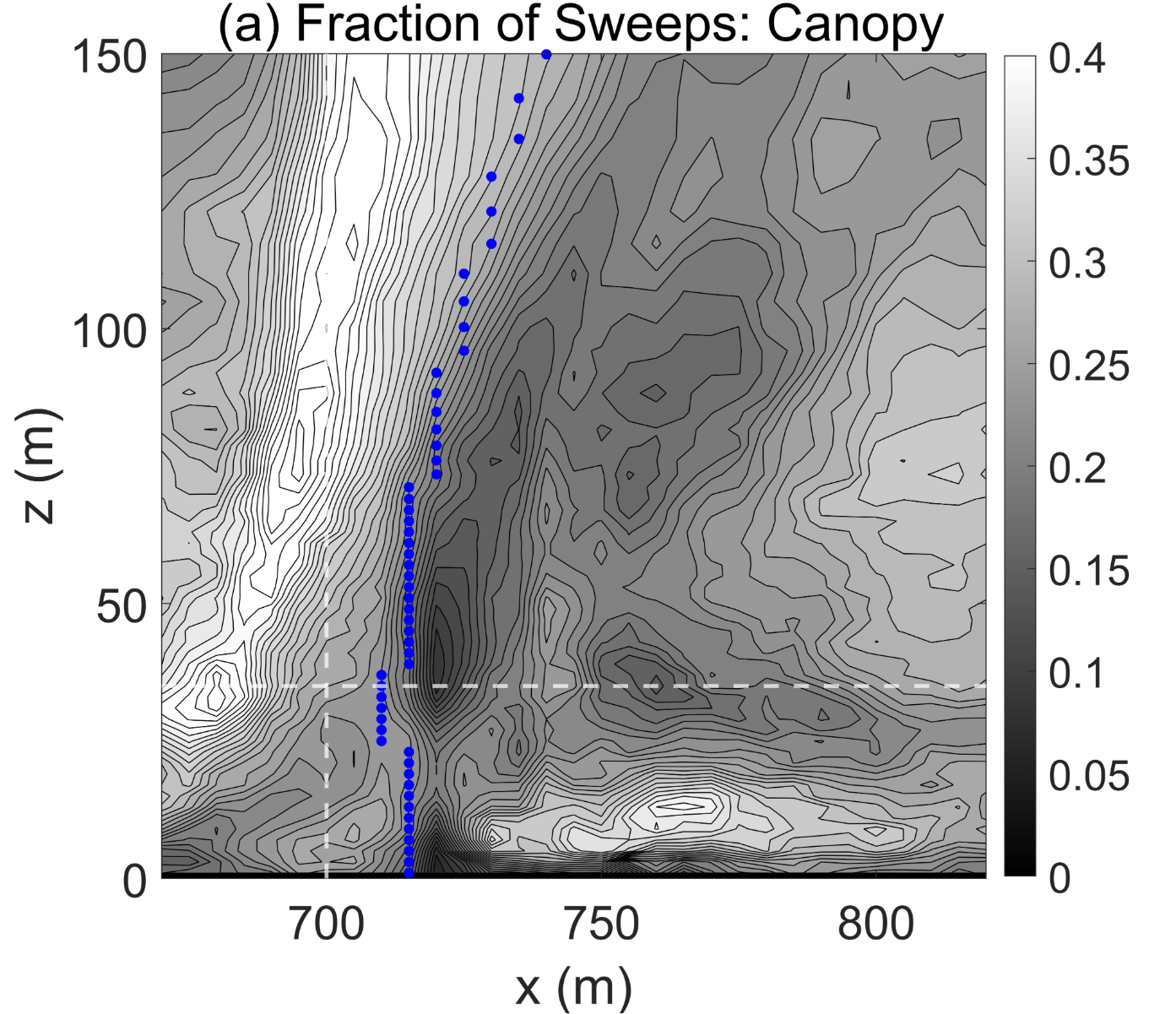}  &
    \includegraphics[scale=0.17]{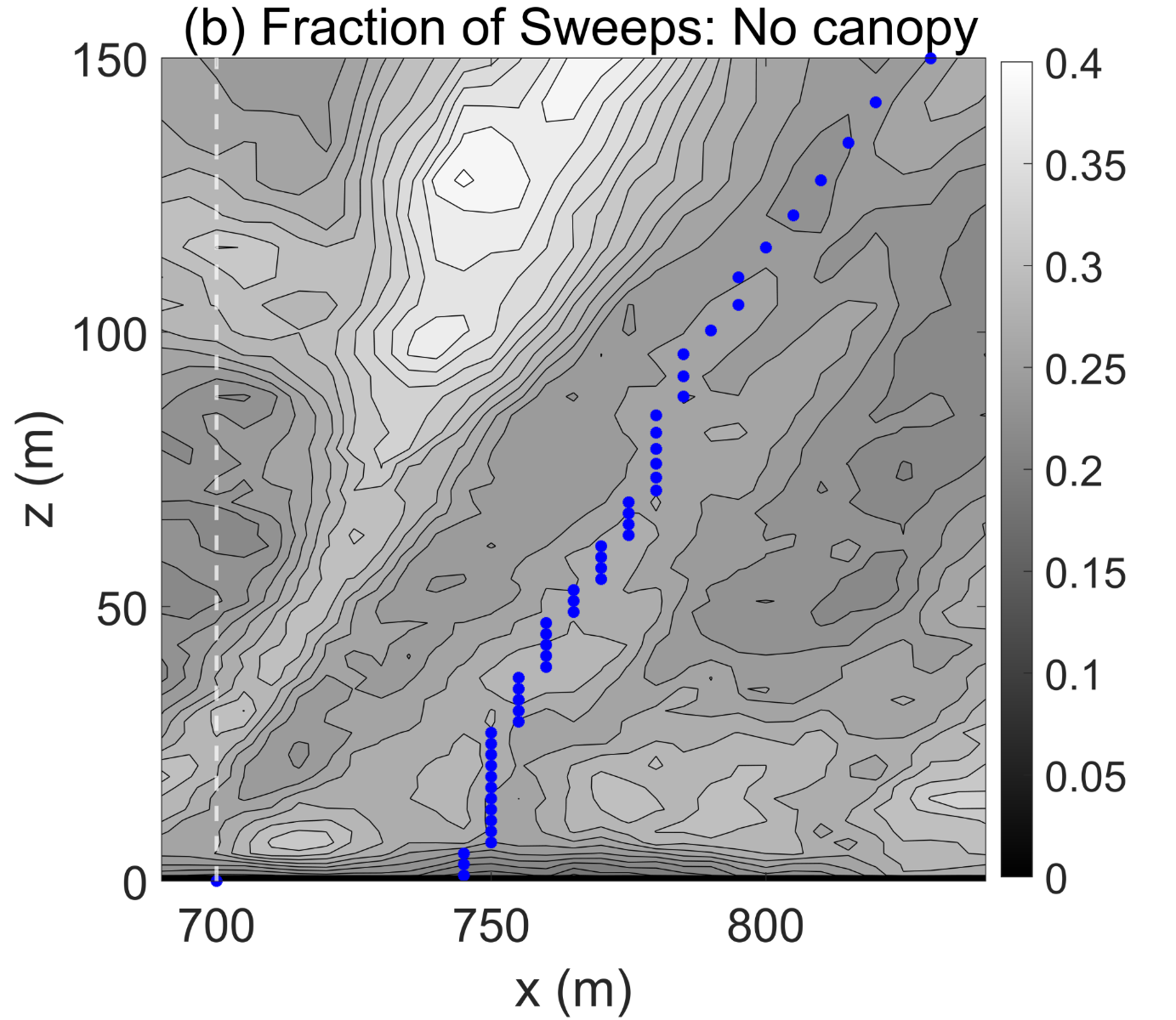} \\
        \includegraphics[scale=0.17]{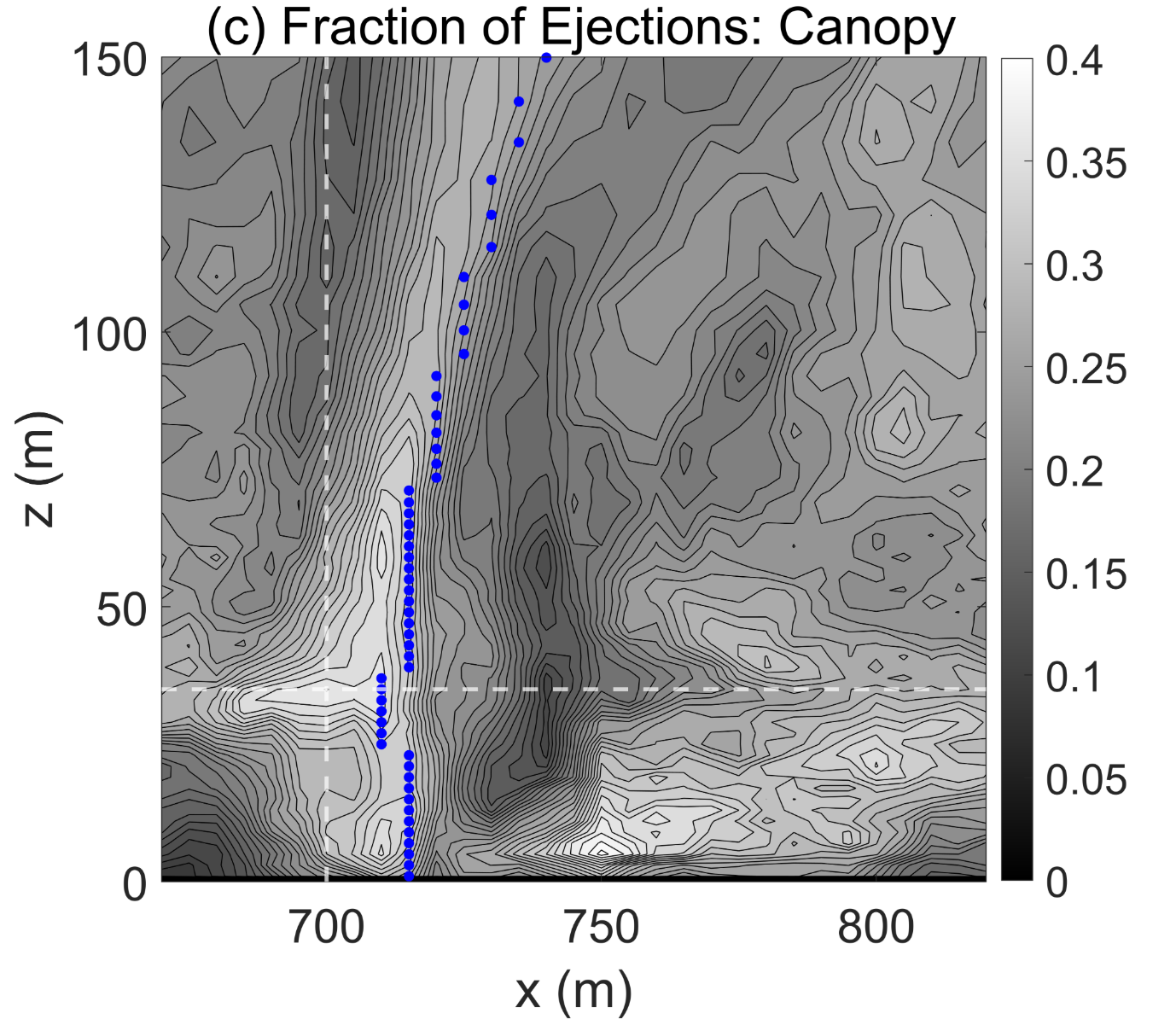} & \includegraphics[scale=0.17]{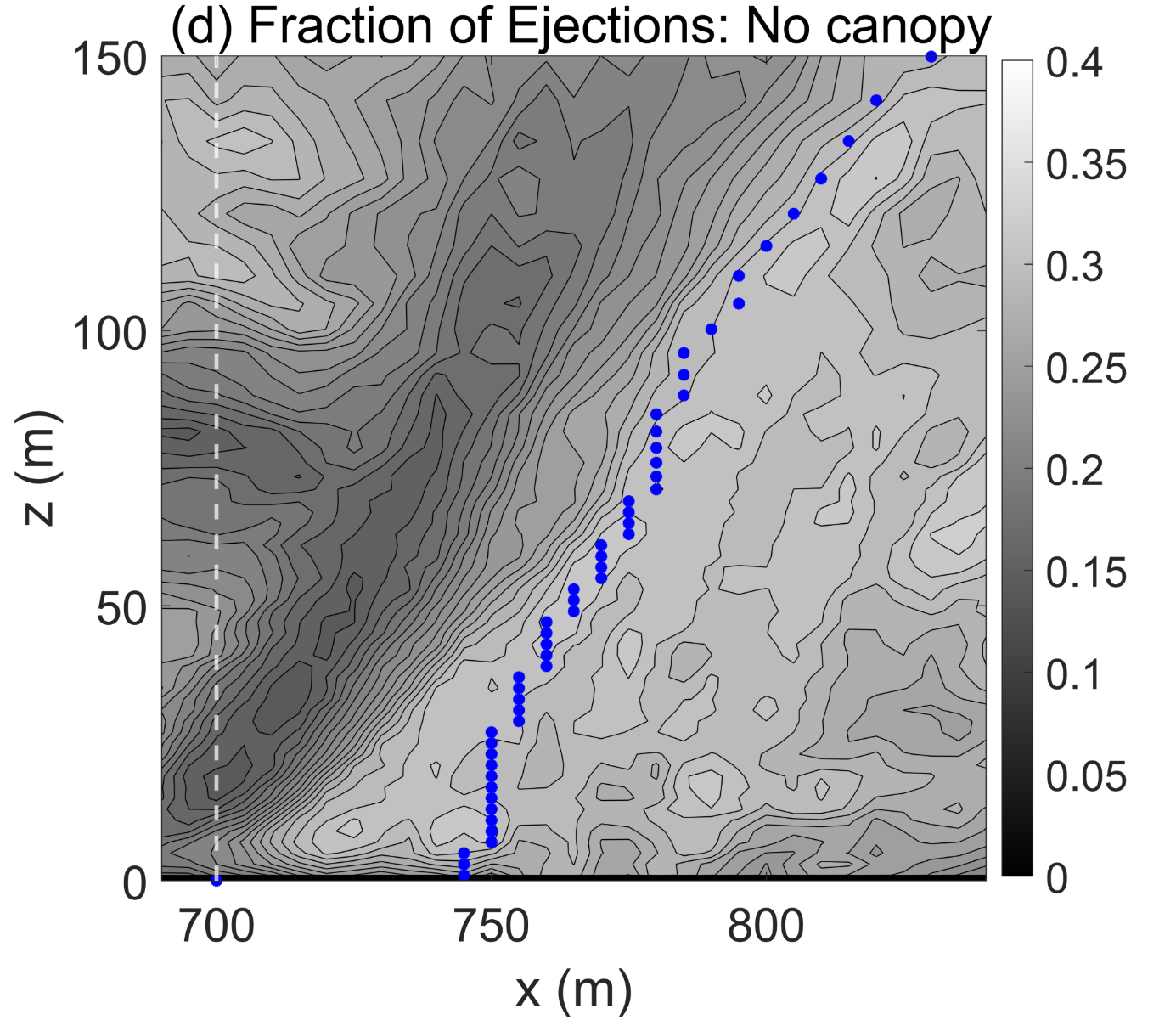}   \\
 \includegraphics[scale=.17]{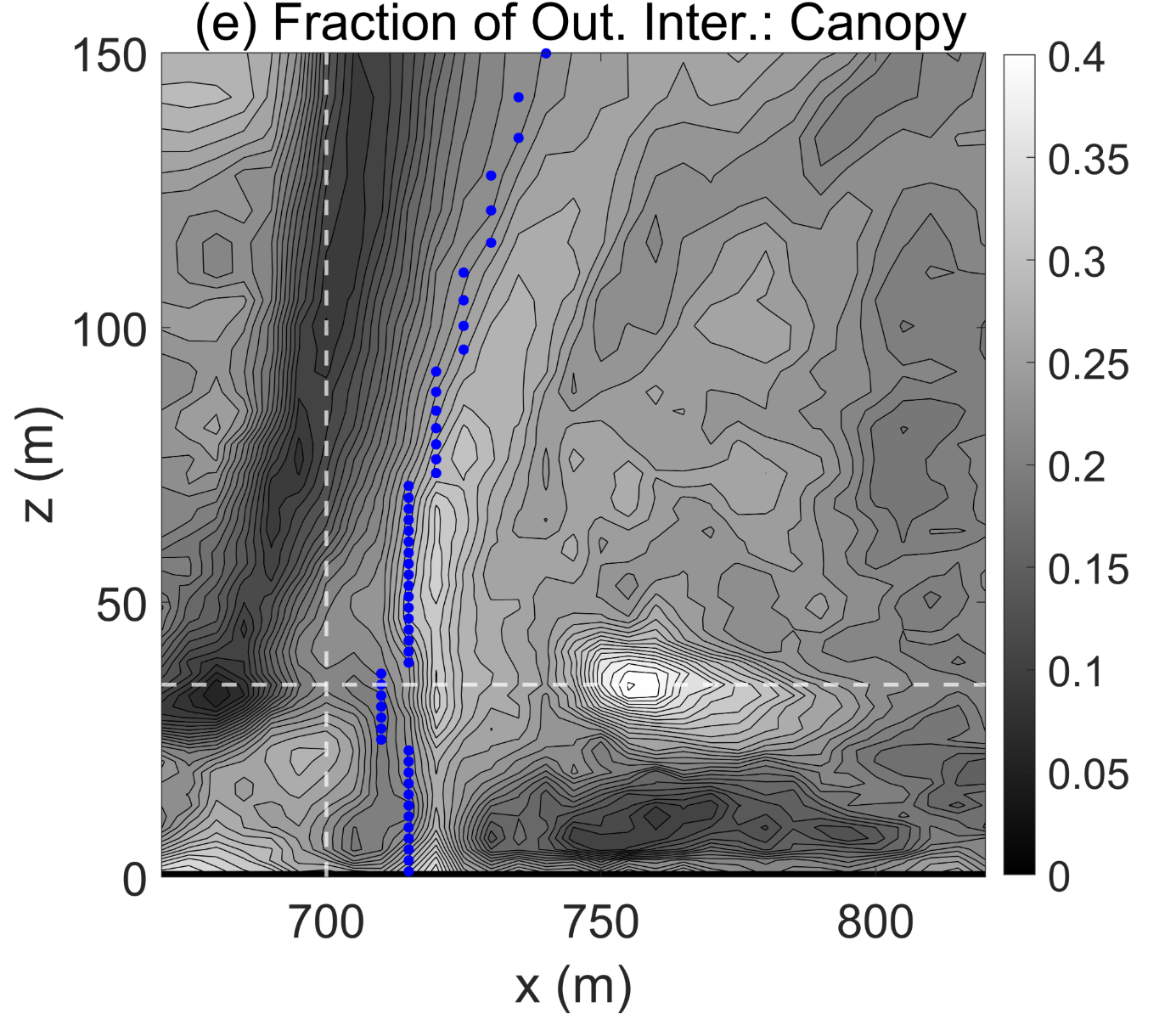} &
  \includegraphics[scale=.17]{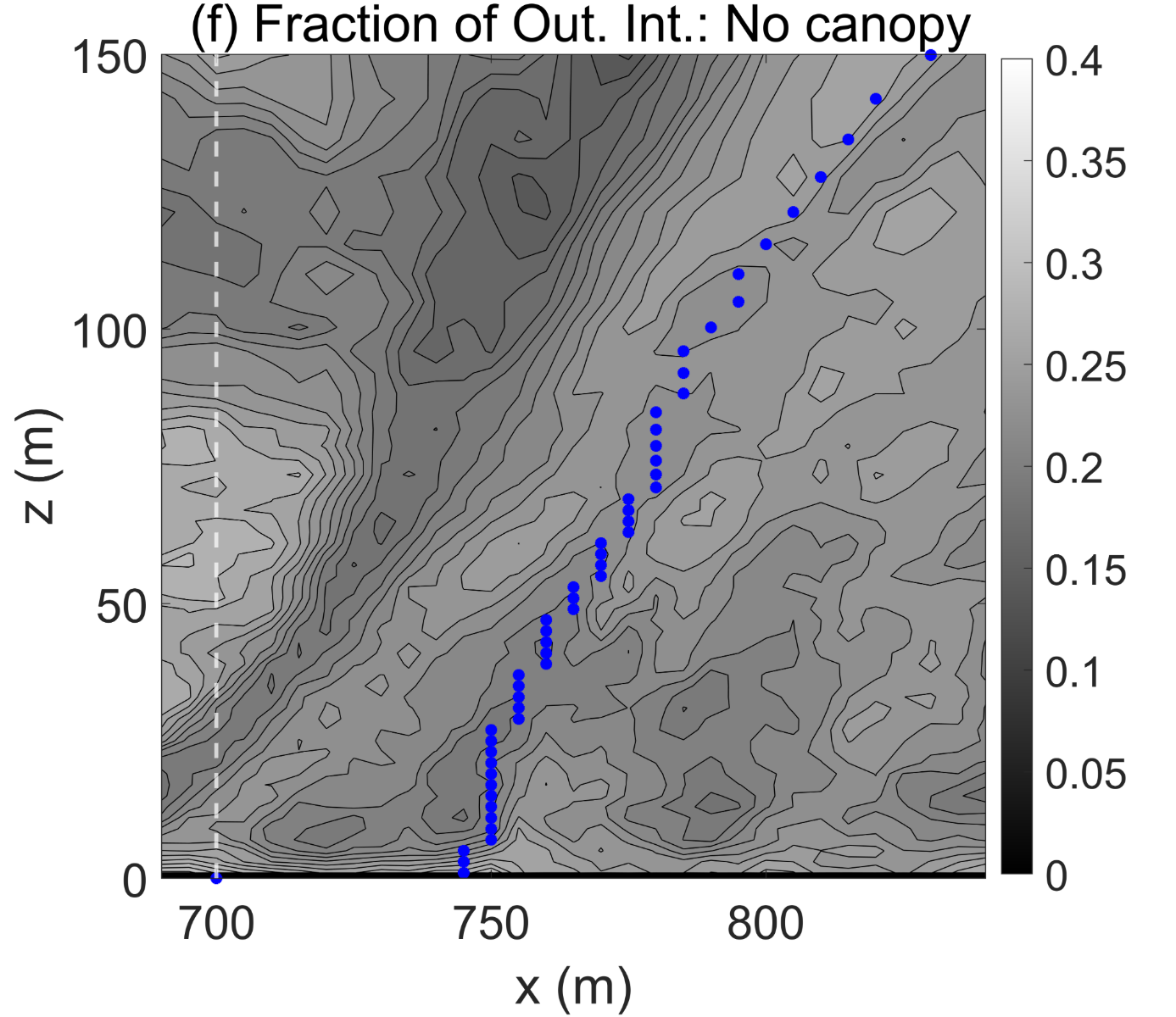} \\
   \includegraphics[scale=.17]{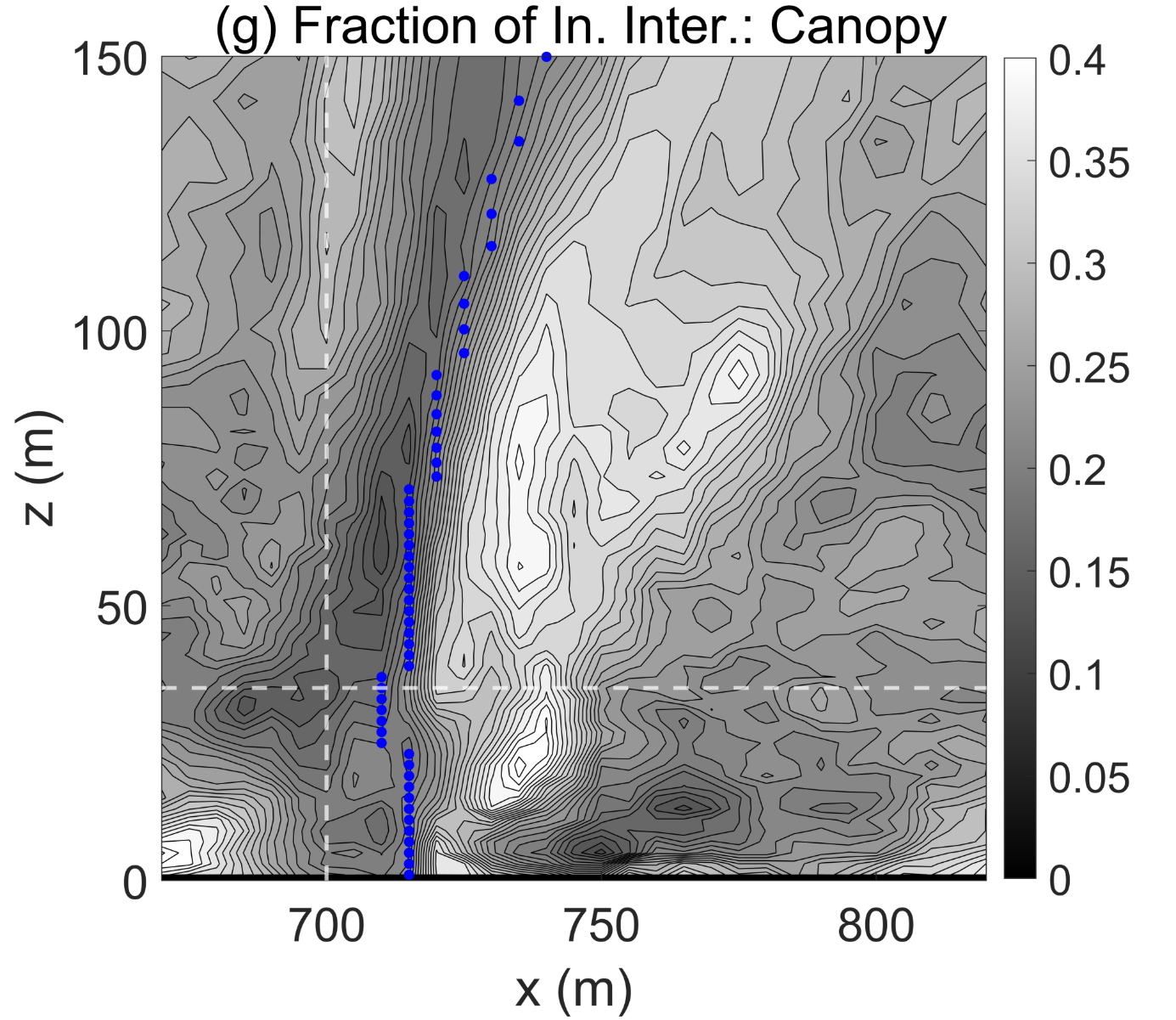} & 
  \includegraphics[scale=.17]{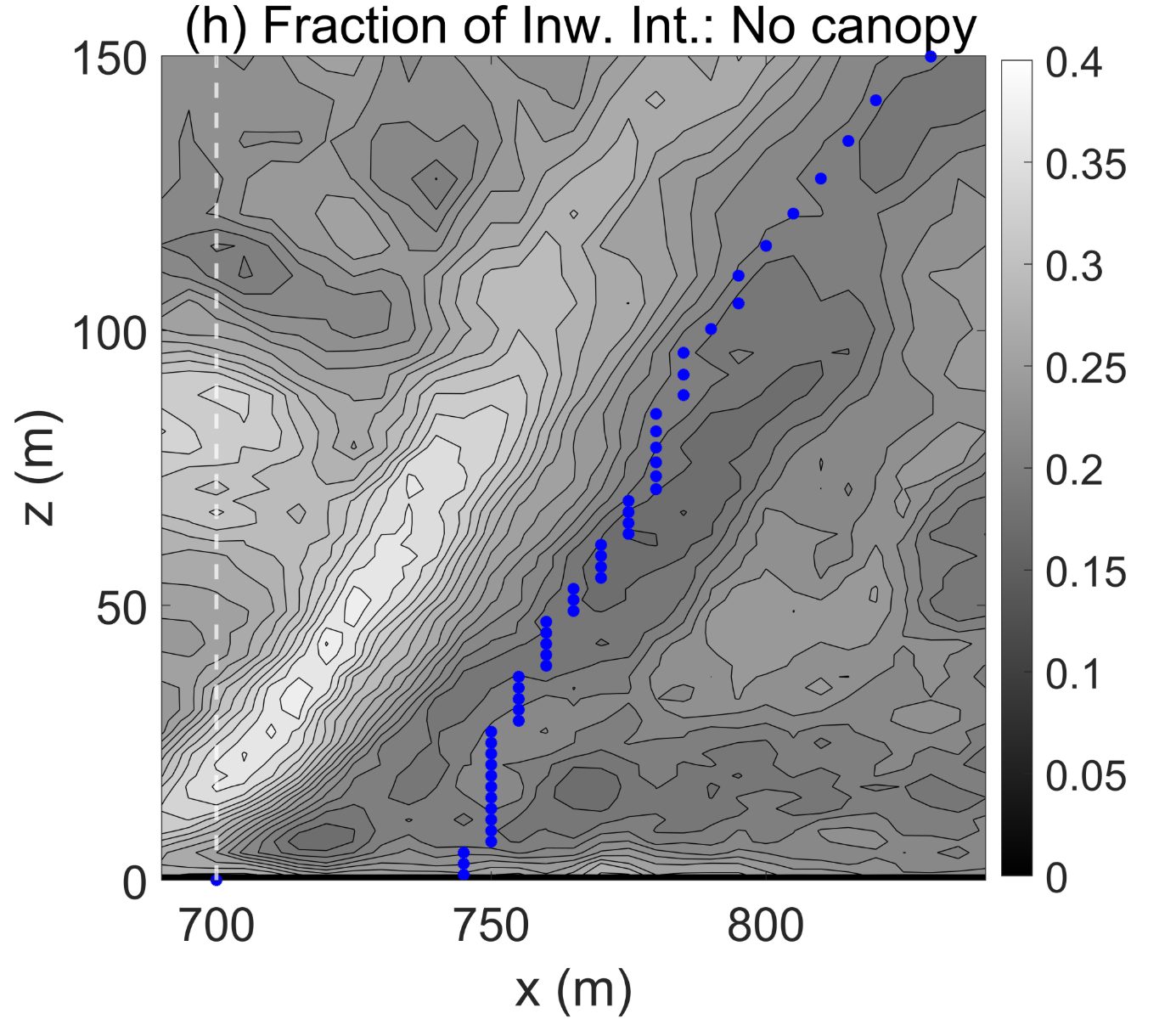} 
    \end{tabular}
    \caption{Fraction of sweeps in the (a) canopy and (b) no-canopy cases; ejections in the (c) canopy and (d) no-canopy cases; outward interactions in the (e) canopy and (f) no-canopy cases; inward interactions in the (g) canopy and (h) no-canopy cases. Blue dots represent the mean plume centerline in Zone 1}
    \label{fig_event_frac}
\end{figure}

Figure~\ref{fig_event_frac} shows the event fractions for each momentum-flux event near the surface/buoyancy source in both, the presence and absence of the canopy. It is observed that in the regions where the ejections dominate the momentum-flux contribution, sweeps dominate in occurrence frequency (compare Figs.~\ref{fig_event_frac}(a)--(b) to Figs.~\ref{fig_uw}(c)--(d)). Moreover, relative to sweeps, the fraction of ejections appears to be relatively suppressed 
(Figs.~\ref{fig_event_frac}(c)--(d)). Similarly, the fraction of inward interactions appears to be higher relative to outward interactions in regions where outward interactions dominate the momentum-flux contribution, in both canopy and no-canopy cases (compare Figs~\ref{fig_event_frac}(g)--(h) to Figs.~\ref{fig_uw}(e)--(f)). Moreover, relative to inward interactions, the fraction of outward interactions is suppressed 
(Figs.~\ref{fig_event_frac}(e)--(h)). In previous studies \cite{heilman2021observations, heilman2021turbulent}, the fraction of sweeps was found to increase at all heights within the canopy as a fire-front passed by a measuring tower. Our analysis suggests an increase in sweep frequencies from the ambient above the canopy down to the canopy top (Fig.~\ref{fig_event_frac}(a)), but not below. Within the canopy, ejections appear to be more frequent in our case (Fig.~\ref{fig_event_frac}(c)), which is inconsistent with experimental data. Moreover, experiments have also demonstrated a higher increase in the occurrence frequency of outward interactions relative to inward interactions within the canopy, which also appears to be inconsistent with our results (Fig.~\ref{fig_event_frac}(e) and (g)). However, the increase in occurrence frequency of inward interactions in the no-canopy environment (Fig.~\ref{fig_event_frac}(h)) is consistent with experiments. Thus, while there is good agreement between our results and experiments for contributions from the momentum-flux events, there are inconsistencies associated with event fractions. These inconsistencies may be attributed to differing canopy densities, wind-forcing conditions (e.g., firelines backing against the wind), heat-source geometries and intensities, or insufficient observations. They highlight the need for more burn experiments and computational simulations over a wide range of forcing conditions and geometrical constraints, especially in canopy environments.

\section{Conclusions and Future Work}\label{sect_conc}
In this study, we have investigated the flow features characteristic of the interaction of a buoyant plume with a mild cross-wind as informed by the presence of a tall, spatially homogeneous plant canopy. This is accomplished using idealized LES with a low-complexity setup sans a combustion model, in which the source of buoyancy, comprising a surface patch of a high sensible heat flux (100 times) relative to the ambient, is static. Apart from attempting to identify the key features distinguishing the effects of the canopy from no-canopy environments, we also address the usefulness of such a setup in replicating fire--atmosphere interaction features as observed in previous studies. 

As the plume interacts with the crosswind, counter-rotating vortex pairs (CVPs) are formed near the source as the flow twists and rotates into the leeward side of the buoyancy source from either side. The rising plume comprises ring-like, spiraling vortical structures representing the entrainment of ambient vorticity into the plume core, akin to processes associated with tornado formation as documented in the literature. The vorticity intensifies at the windward edge, where the ring-like structures are compressed together, and tilted “hair-pin”-like vortical structures exist on the leeward side. Flow patterns aloft exhibit helical motions as the CVPs aloft propagate downstream, trailing the plume. To explore the effect of changing the source geometry, we examine the flow field surrounding a high-aspect ratio (25) source resembling a line source of buoyancy. Tower-trough-like structures are found to exist, with forward burst-like motions in the trough regions, which potentially carry heat downstream as they penetrate the canopy. These flow features show promising consistencies with prior studies on buoyant plumes in cross-flow without canopies, propagating fires in laboratory-scale fuel beds, static fires in field-scale experiments, and FIRETEC simulations. This suggests that this simplified setup can be promising for capturing much of the plume dynamics, while circumventing the computational overhead associated with the complex system of equations governing fire spread in models and relieving the pressure on the resources deployed during controlled burns to collect extensive measurements.

Differences are observed in the plume behavior between the simulated canopy and no-canopy environments, both near the buoyancy source and in the far-field region. Near the source, the plume tilts relatively less steeply in the canopy case due to the canopy drag on the plume. Moreover, on the leeward side of the mean plume centerline, a recirculation region is formed within the canopy as air is entrained into the plume from there. As the flow approaches the region of increased pressure upstream of the recirculation, it decelerates and is entrained into the plume. In this manner, the recirculation zone obstructs the upstream flow, and two distinct regions are formed on either side of the plume centerline, with the leeward side being sheltered from the upstream flow. In contrast, no recirculation region exists in the no-canopy case, and entrainment from the upstream side of the plume centerline is stronger. The flow impinges upon the rising plume from the upstream side, and the plume tilts more steeply. The plume transitions from the rise phase to the far-field at a higher altitude in the canopy case, indicating the existence of a canopy-induced scale encompassing its aerodynamic effects on the rising plume. Moreover, the spatially oscillatory behavior of the far-field mean plume centerline is subdued compared to the no-canopy case.

The spatial distribution of turbulent momentum flux events near the buoyancy source also differs between the two environments. Above the canopy, there is downward transfer of momentum predominantly via ejections (upward fluxes of low momentum) toward the canopy top upstream of the mean plume centerline; however, sweeps (downward fluxes of high momentum) prevail within the canopy volume, upstream of the mean plume centerline. On the leeward side, sweeps dominate ejections. However, counter-gradient motions play a more significant role in transferring momentum away from the buoyancy source, with outward interactions (upward fluxes of high momentum) being most dominant both within and above the canopy. These observations are consistent with measurements collected within tall canopies during prescribed burns. In contrast, the no-canopy environment shows a different pattern: counter-gradient motions near the surface are sandwiched between an ejection-dominated region on the upstream side and a sweep-dominated region on the downstream side. On the upstream side of the mean plume centerline, outward interactions compete with ejections--again, an observation consistent with grassland burns. The physics-based insights, obtained from our analysis, into the differences in plume behavior between canopy and no-canopy environments can be validated against experimental data and can be utilized for the development of improved parameterizations within fire- and plume-behavior models.

Notwithstanding the promise of this setup and analysis, there are limitations that need to be addressed. While our results regarding the contributions from the momentum-flux events agree with those from experiments, the inferences regarding event fractions are inconsistent with observations. These inconsistencies may be attributed to differing canopy densities, wind-forcing conditions (e.g., firelines backing against the wind), heat-source geometries and intensities, or insufficient observations. Additional computational simulations can be aimed at exploring the effects of a range of wind and buoyancy forcing conditions, source geometry, and canopy characteristics.  
The effect of the domain parameters and the fetch over which the flow at the inlet adjusts to the canopy presence before encountering the plume may need to be investigated to ensure the independence of the results from these effects. Moreover, for a comparison with coherent structures observed in field-scale burns within tall canopies, the temporal resolution may need to be adjusted to match the higher sampling frequency from experiments. Future work can also involve a comparison of the steady-state features from the current analysis with the transient dynamics of propagating experimental fires in the field. Despite its limitations, the analysis advances our knowledge regarding plume--cross-wind interactions in the presence of a canopy, which remains sparse in the literature. Moreover, it lays the foundation for the development of scaling laws characterizing plume trajectories in canopy environments, which, in turn, informs the development of
improved predictive models for plume behavior.
\section*{Acknowledgments}
 Banerjee acknowledges the funding support from the University of California Office of the President (UCOP) grant LFR-20-653572 (UC Lab-Fees); the National Science Foundation (NSF) grants NSF-AGS-PDM-2146520 (CAREER), NSF-OISE-2114740 (AccelNet), NSF-CPS-2209695, NSF‐ECO‐CBET‐2318718, NSF-DMS-
2335847, and NSF-RISE-2536815; the United States Department of Agriculture (USDA) grant 2021-67022-35908 (NIFA); and a cost reimbursable agreement with the USDA Forest Service 20-CR-11242306-072. Desai acknowledges the Graduate Dean's Dissertation Fellowship, the Distinguished Public Impact Fellowship, and the Henry Samueli Endowed Fellowship at UC Irvine. Desai acknowledges that part of the work involved in writing the manuscript was done under the auspices of the Lawrence Livermore National Laboratory (LLNL): the Release Number is \textbf{LLNL-JRNL-2012493}. 

\section*{Data Availability Statement}
We used the PALM model (Maronga et al. 2015 \cite{maronga2015parallelized}, Maronga et al. 2020 \cite{ maronga2020overview}), which can be downloaded from its official site: \url{https://palm.muk.uni-hannover.de/trac/wiki/doc/install}. The parameter file and the Python script to generate the NETCDF (static driver) file with canopy geometry and surface heat flux information, which were modified for our analysis, can be found at \url{https://palm.muk.uni-hannover.de/trac/wiki/doc/app/plant_canopy_parameters}. 
 
\bibliographystyle{ieeetr}
\bibliography{references3}
\end{document}